\documentclass[aps,prl,twocolumn,superscriptaddress,showpacs,floatfix,longbibliography,nofootinbib,preprintnumbers]{revtex4-2}
\usepackage[utf8]{inputenc}
\usepackage{graphicx}
\usepackage{color}  
\usepackage{amsmath}
\usepackage{dcolumn}
\usepackage[english]{babel}
\usepackage[colorlinks=true,linkcolor=blue,citecolor=blue, urlcolor=blue]{hyperref}

\newcommand{\bra}[1]{\langle #1 \vert}
\newcommand{\ket}[1]{\vert #1 \rangle}
\newcommand{\scal}[2]{\langle #1 \vert #2 \rangle}
\newcommand{\elma}[3]{\bra{#1} #2 \ket{#3}}

\newcommand{\nene}{$^{20}$Ne$^{20}$Ne}
\newcommand{\oooo}{$^{16}$O$^{16}$O}
\newcommand{\pbpb}{$^{208}$Pb$^{208}$Pb}
\newcommand{\xexe}{$^{129}$Xe$^{129}$Xe}

\newcommand{\uuuu}{$^{238}$U$^{238}$U}

\newcommand{\oo}{$^{16}$O}
\newcommand{\nee}{$^{20}$Ne}

\newcommand{\mpt}{$\langle p_T \rangle$}

\definecolor{amber}{rgb}{1.0, 0.49, 0.0}

\begin{document}

\title{The unexpected uses of a bowling pin: exploiting $^{20}$Ne isotopes for precision characterizations of collectivity in small systems}




\author{Giuliano Giacalone}
\email{giacalone@thphys.uni-heidelberg.de}
\affiliation{Institut f\"{u}r Theoretische Physik, Universit\"{a}t Heidelberg, Philosophenweg 16, 69120 Heidelberg, Germany}

\author{Benjamin Bally}
\affiliation{ESNT, IRFU, CEA, Universit\'e Paris-Saclay, 91191 Gif-sur-Yvette, France}

\author{Govert Nijs}
\affiliation{Theoretical Physics Department, CERN, CH-1211 Gen\`eve 23, Switzerland}

\author{Shihang Shen}
\affiliation{Institute for Advanced Simulation and Institut f\"ur Kernphysik, Forschungszentrum J\"ulich, D-52425 J\"ulich, Germany}

\author{\\Thomas Duguet}
\affiliation{IRFU, CEA, Universit\'e Paris-Saclay, 91191 Gif-sur-Yvette, France}
\affiliation{KU Leuven, Instituut voor Kern- en Stralingsfysica, 3001 Leuven, Belgium}

\author{Jean-Paul Ebran}
\affiliation{CEA, DAM, DIF, 91297 Arpajon, France}
\affiliation{Universit\'e Paris-Saclay, CEA, Laboratoire Mati\`ere en Conditions Extr\^emes, 91680 Bruy\`eres-le-Ch\^atel, France}

\author{Serdar Elhatisari}
\affiliation{Faculty of Natural Sciences and Engineering, Gaziantep Islam Science and Technology University, Gaziantep 27010, Turkey}
\affiliation{Helmholtz-Institut f\"ur Strahlen- und Kernphysik and Bethe Center for Theoretical Physics, Universit\"at Bonn, D-53115 Bonn, Germany}

\author{Mikael Frosini}
\affiliation{CEA, DES, IRESNE, DER, SPRC, LEPh, 13115 Saint-Paul-lez-Durance, France}

\author{Timo A. L\"ahde}
\affiliation{Institut f\"ur Kernphysik, Institute for Advanced Simulation and J\"ulich Center for Hadron Physics, Forschungszentrum Julich, D-52425 J\"ulich, Germany}
\affiliation{Center for Advanced Simulation and Analytics (CASA), Forschungszentrum Julich, D-52425 J\"ulich, Germany}

\author{\\Dean Lee}
\affiliation{Facility for Rare Isotope Beams and Department of Physics and Astronomy, Michigan State University, MI 48824, USA}

\author{Bing-Nan Lu}
\affiliation{Graduate School of China Academy of Engineering Physics, Beijing 100193, China}

\author{Yuan-Zhuo Ma}
\affiliation{Facility for Rare Isotope Beams and Department of Physics and Astronomy, Michigan State University, MI 48824, USA}

\author{Ulf-G. Mei\ss ner}
\affiliation{Helmholtz-Institut f\"ur Strahlen- und Kernphysik and Bethe Center for Theoretical Physics, Universit\"at Bonn, D-53115 Bonn, Germany}
\affiliation{Institute for Advanced Simulation, Institut f\"ur Kernphysik, and Julich Center for Hadron Physics, Forschungszentrum J\"ulich, D-52425 J\"ulich, Germany}
\affiliation{Tbilisi State University, 0186 Tbilisi, Georgia}

\author{Jacquelyn Noronha-Hostler}
\affiliation{Illinois Center for Advanced Studies of the Universe, Department of Physics, University of Illinois at
Urbana-Champaign, Urbana, IL 61801, USA}

\author{\\Christopher Plumberg}
\affiliation{Natural Science Division, Pepperdine University, Malibu, CA 90263, USA}

\author{Tom\'as R.\ Rodr\'iguez}
\affiliation{Departamento de Estructura de la Materia, F\'isica T\'ermica y Electr\'onica and IPARCOS, Universidad Complutense de Madrid, E-28040 Madrid, Spain}

\author{Robert Roth}
\affiliation{Institut f\"ur Kernphysik, Technische Universit\"at Darmstadt, 64289 Darmstadt, Germany}
\affiliation{Helmholtz Forschungsakademie Hessen f\"ur FAIR, GSI Helmholtzzentrum, 64289 Darmstadt, Germany}

\author{Wilke van der Schee}
\affiliation{Theoretical Physics Department, CERN, CH-1211 Gen\`eve 23, Switzerland}
\affiliation{Institute for Theoretical Physics, Utrecht University, 3584 CC Utrecht, The Netherlands}
\affiliation{NIKHEF, Amsterdam, The Netherlands}

\author{Vittorio Som\`a}
\affiliation{IRFU, CEA, Universit\'e Paris-Saclay, 91191 Gif-sur-Yvette, France}

\begin{abstract}
Whether or not femto-scale droplets of quark-gluon plasma (QGP) are formed in so-called \textit{small systems} at high-energy colliders is a pressing question in the phenomenology of the strong interaction. For proton-proton or proton-nucleus collisions the answer is inconclusive due to the large theoretical uncertainties plaguing the description of these processes. While upcoming data on collisions of $^{16}$O nuclei may mitigate these uncertainties in the near future, here we demonstrate the unique possibilities offered by complementing \oooo{} data with collisions of $^{20}$Ne ions.  We couple both NLEFT and PGCM \textit{ab initio} descriptions of the structure of $^{20}$Ne and $^{16}$O to hydrodynamic simulations of \oooo{} and \nene{} collisions at high energy. We isolate the imprints of the bowling-pin shape of $^{20}$Ne on the collective flow of hadrons, which can be used to perform quantitative tests of the hydrodynamic QGP paradigm. In particular, we predict that the elliptic flow of \nene{} collisions is enhanced by as much as 1.170(8)$_{\rm stat.}$(30)$_{\rm syst.}$ for NLEFT and 1.139(6)$_{\rm stat.}$(39)$_{\rm syst.}$ for PGCM relative to \oooo{} collisions for the 1\% most central events. At the same time, theoretical uncertainties largely cancel when studying relative variations of observables between two systems. This demonstrates a method based on experiments with two light-ion species for precision characterizations of the collective dynamics and its emergence in a small system.
\end{abstract}

\preprint{CERN-TH-2024-021}

\maketitle

\paragraph{Introduction.} 

A central motivation for the program of ultra-relativistic nuclear collisions is to access bulk properties of QCD matter that emerge in conditions similar to those found in the early Universe or in extreme astrophysical objects \cite{Busza:2018rrf}\@. A prime example is the quark-gluon plasma (QGP), the hot phase of QCD matter that behaves like a near-perfect fluid \cite{Teaney:2009qa}\@. Hydrodynamic behavior is inferred from the harmonic spectrum of the azimuthal distributions of final-state hadrons \cite{Ollitrault:2023wjk},
\[
\frac{dN_{\rm ch}}{d\eta\,d^2 {\bf p} } = \frac{dN_{\rm ch}}{d\eta\,d p_T} \frac{1}{2\pi} \left (1 + 2 \sum_{n=1}^{\infty} v_n \cos n(\phi-\phi_n) \right ),
\]
where $dN_{\rm ch}/d\eta\,d^2{\bf p}$ is the charged hadron distribution differential in pseudorapidity, $\eta$, and transverse momentum, ${\bf p}$, with $p_T=|{\bf p}|$ and $\phi$ the azimuthal angle. The coefficients $v_n$ quantify the \textit{anisotropic flow}\@. In hydrodynamics, $v_n$ arise as a response of the system created in the interaction region to the anisotropy of its geometry, as dictated by an emergent pressure-gradient force \cite{Ollitrault:1992bk}, the hallmark of hydrodynamic behavior. An elliptical deformation of the interaction region leads to elliptic flow, $v_2$, a triangular deformation to $v_3$, and so on \cite{Teaney:2010vd}\@. 

Observations of anisotropic flow in \textit{small systems} \cite{Nagle:2018nvi,Schenke:2021mxx}, such as proton-nucleus and proton-proton collisions, have triggered tremendous efforts investigating whether a QGP description is appropriate even in regimes where applying hydrodynamics becomes hard to justify \cite{Noronha:2024dtq, Rocha:2023ilf}\@. Theoretical studies have either pushed hydrodynamic simulations to extreme situations \cite{Weller:2017tsr,Mantysaari:2017cni,Moreland:2018gsh,Schenke:2019pmk,Zhao:2020pty,YuanyuanWang:2023meu}, analyzed in detail the transition from kinetic theory to hydrodynamics \cite{Kurkela:2018ygx,Kurkela:2018qeb,Kurkela:2018wud,Kurkela:2019kip,Kurkela:2020wwb,Ambrus:2021fej,Ambrus:2022koq,Ambrus:2022qya}, or studied the emergence of collectivity via other mechanisms \cite{Altinoluk:2020wpf}\@. Small systems pose, thus, a fundamental challenge rooted in the issue of the thermalization and hydrodynamization of QCD matter \cite{Romatschke:2017ejr,Blaizot:2019scw,Schlichting:2019abc,Berges:2020fwq,Soloviev:2021lhs}\@. 

To advance our knowledge of small systems, one has to isolate in the experimental data information able to discriminate theoretical approaches. A breakthrough in this direction would be the identification of a correlation between the final-state anisotropy in momentum space ($v_n$) and the deformation of the initial-state geometry, supporting an underlying hydrodynamic-type scenario. This strategy has been pursued at the Relativistic Heavy Ion Collider (RHIC) in a system-geometry scan comparing $p{}^{197}$Au and $d{}^{197}$Au collisions at the same beam energy. Defining $v_2\{2\}\equiv\sqrt{ \langle v_2^2 \rangle}$ as the elliptic flow at a given multiplicity, both the PHENIX collaboration \cite{PHENIX:2018lia} and the STAR collaboration \cite{STAR:2022pfn,STAR:2023wmd} observe
\[
v_2\{2\}_{d{\rm ^{197}Au}} >  v_2\{2\}_{p{\rm ^{197}Au}}.
\]
This constitutes a plausible signature of the elliptical geometry of the system formed when a deuterium impinges onto a large gold target. Similarly, at the Large Hadron Collider (LHC) one observes \cite{CMS:2013jlh,ATLAS:2017hap,ALICE:2019zfl}
\begin{align*}
v_2\{2\}_{{\rm ^{208}Pb}{\rm ^{208}Pb}} &>  v_2\{2\}_{p{\rm ^{208}Pb}}, \\
\hspace{15pt}v_3\{2\}_{{\rm ^{208}Pb}{\rm ^{208}Pb}} &\approx  v_3\{2\}_{p{\rm ^{208}Pb}}.
\end{align*}
The enhancement of elliptic flow in \pbpb{} collisions is interpreted as coming from the intrinsic ellipticity of the overlap area for off-central collisions, i.e.~collisions that are not head-on. These observations hint at the role played by the collision geometry, but employ proton-nucleus collisions as a baseline of a system that does not present any intrinsic shape. This presents two drawbacks. First, proton-nucleus collisions have a different longitudinal structure than nucleus-nucleus collisions (including $d{}^{197}$Au collisions \cite{PHENIX:2021ubk,Zhao:2022ugy}), whose geometry is better correlated across rapidities \cite{CMS:2015xmx,Bozek:2015bna}\@. Second, the geometry of proton-nucleus collisions largely depends on the proton structure at low values of the Bjorken $x$ variable, which is poorly understood \cite{Schenke:2021mxx}\@. Thus, it would be desirable to isolate signatures of the geometry of the initial states in the scattering of actual ions, presenting a well-defined notion of an interaction region.

Upcoming data on collisions of $^{16}$O isotopes is expected to mitigate these issues \cite{Brewer:2021kiv}\@. Preliminary data from \oooo{} collisions were recently presented at the Quark Matter 2023 conference by the STAR collaboration \cite{Huang:2023viw}\@. At the CERN LHC, a run of \oooo{} collisions is expected to take place in 2025\@. Comparing peripheral \pbpb{} (or \xexe{}) collisions and central \oooo{} collisions should reveal \cite{Schenke:2020mbo,Nijs:2021clz}:
\[
v_2\{2\}_{\rm ^{208}Pb^{208}Pb} > v_2\{2\}_{\rm ^{16}O^{16}O}.
\]
However, comparing highly peripheral \pbpb{} collisions with central \oooo{} collisions is suboptimal, as it does not resolve issues related to the definition of an overlap area and the longitudinal structure.

In this Letter, we demonstrate an alternative more robust approach to isolate the impact of the initial-state geometry. We study central collisions of light ions presenting different shapes. Differences in the collective flow between two collision systems would demonstrate the influence of the nuclear geometry, a technique akin to that used to infer nuclear deformation effects in isobar collisions at RHIC \cite{Giacalone:2021uhj,STAR:2021mii,Zhang:2021kxj,Nijs:2021kvn}\@. The advantage with light species is that we benefit from an advanced knowledge of their geometries coming from \textit{ab initio} calculations of nuclear structure \cite{Hergert:2020bxy,Ekstrom:2022yea}\@. The drawback with light-ion collisions is instead that the anisotropy induced by nuclear shapes is only a small correction to the anisotropy induced by large density fluctuations caused by the small numbers of participant nucleons. In other words, with light ions extreme nuclear shapes are required for their fingerprints to be detectable in the final state.

Here, we overcome this issue. We exploit the fact that the stable isotope presenting the most extreme ground-state geometry in the Segr\`e chart, namely $^{20}$Ne, is close in mass to $^{16}$O\@. We argue that having \nene{} data in conjunction with \oooo{} data leads to the observation of unambiguous imprints of the initial-state geometry on the collective flow. This in turn enables one to perform quantitative tests of hydrodynamics in a small system.

\paragraph{Nuclear structure inputs.}
Modern \emph{ab initio} approaches to the nuclear many-body problem aim at solving as exactly as possible Schr\"odinger's equation for nuclear Hamiltonians constructed through chiral effective field theories of low-energy QCD\@. Such approaches are routinely used to describe the structure of light- and medium-mass nuclei \cite{Hagen:2013nca,Hergert:2015awm,Soma:2019bso,Stroberg:2019bch,Tichai:2020dna,Arthuis:2024mnl} and first applications to $^{208}$Pb were even recently reported \cite{Hu:2021trw,Arthuis:2024mnl}\@. In this work, we employ results for the structure of $^{16}$O and $^{20}$Ne derived within the the framework of Nuclear Lattice Effective Field Theory (NLEFT) simulations and the \textit{ab initio} Projected Generator Coordinate Method (PGCM)\@.

The NLEFT framework \cite{Lee:2008fa,Lahde:2019npb,Lee:2020meg} combines the principles of effective field theory with lattice Monte Carlo methods, and is well suited to probe clustering and other collective phenomena in the ground states of nuclei \cite{Elhatisari:2017eno}\@. NLEFT simulations implement a Euclidean time evolution coupled with auxiliary-field Monte Carlo simulations to produce ground-state configurations of nucleons for each realization of the nuclear wave function. The pin-hole algorithm \cite{Elhatisari:2017eno} enables one to keep track of the positions of the nucleons during the Euclidean time evolution while preserving the information about their center-of-mass. The produced nuclear configurations carry, thus, many-body correlations to all orders as dictated by the ground state of the Hamiltonian. We employ a minimal pion-less EFT Hamiltonian with a periodic lattice of eight sites with spacing $a=1.315$\,fm \cite{Lu:2018bat}, which successfully reproduces measured binding energies and charge radii for the isotopes under study. For $^{16}$O, the pinhole configurations are taken from Ref.~\cite{Summerfield:2021oex}, while a new set is calculated for $^{20}$Ne. Due to the larger mass number, these configurations contain a larger fraction of nuclei with a non-unique center-of-mass due to the periodicity, as well as a higher number of negative-weight states \cite{Elhatisari:2017eno,Lahde:2019npb} than the $^{16}$O ones. These issues are addressed in the evaluation of our uncertainties for the subsequent hydrodynamic study (see the Supplemental Material (SM))\@. Lastly, we distribute nucleons at each lattice site uniformly between $-a/2$ and $a/2$ while maintaining a minimum inter-nucleon distance, $d_{\rm min}$, to mimic the effect of short-range repulsion.

\begin{figure}[t]
    \centering
    \includegraphics[width=.8\columnwidth]{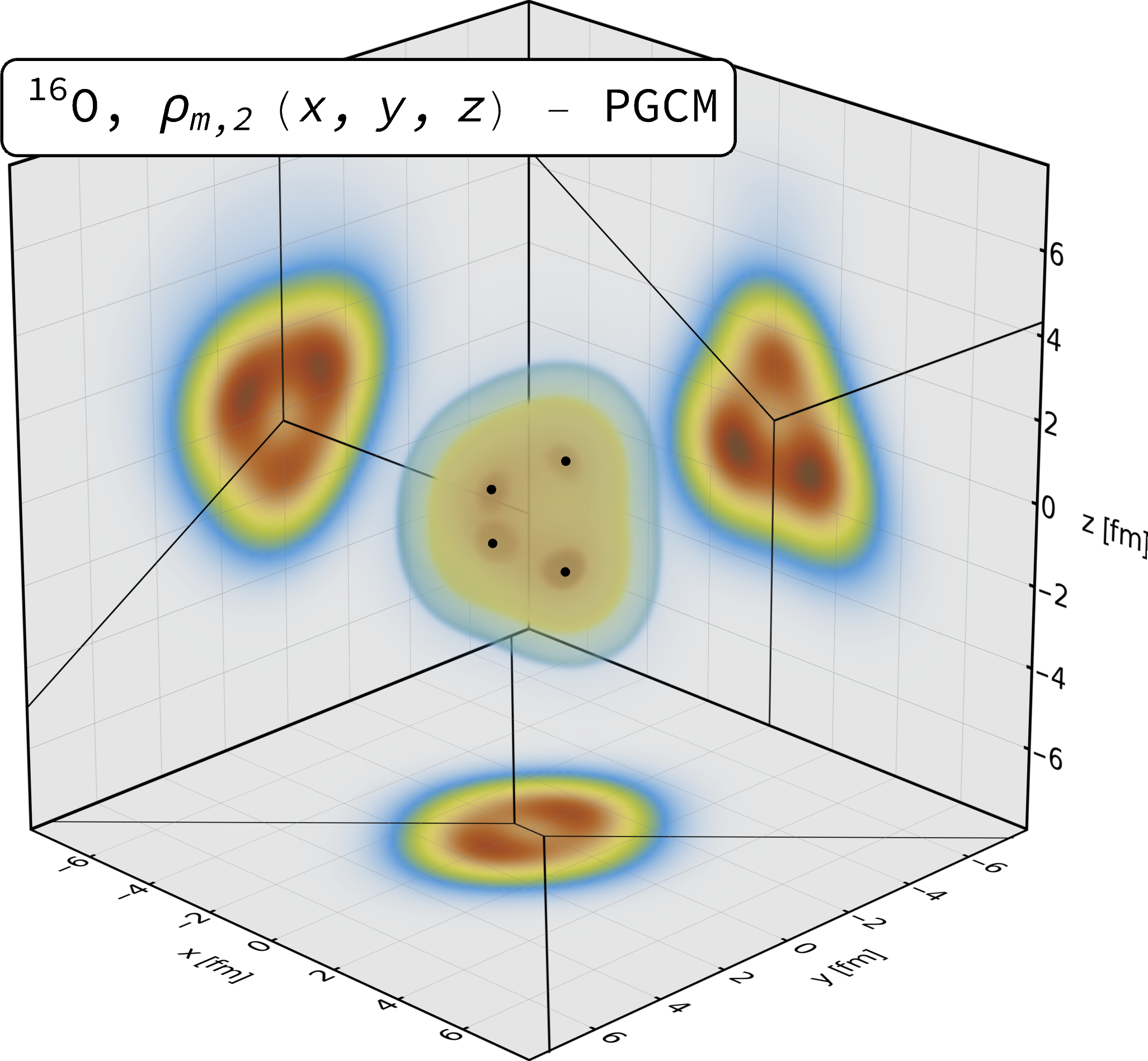}\\
    \vspace{10pt}
    \includegraphics[width=.8\columnwidth]{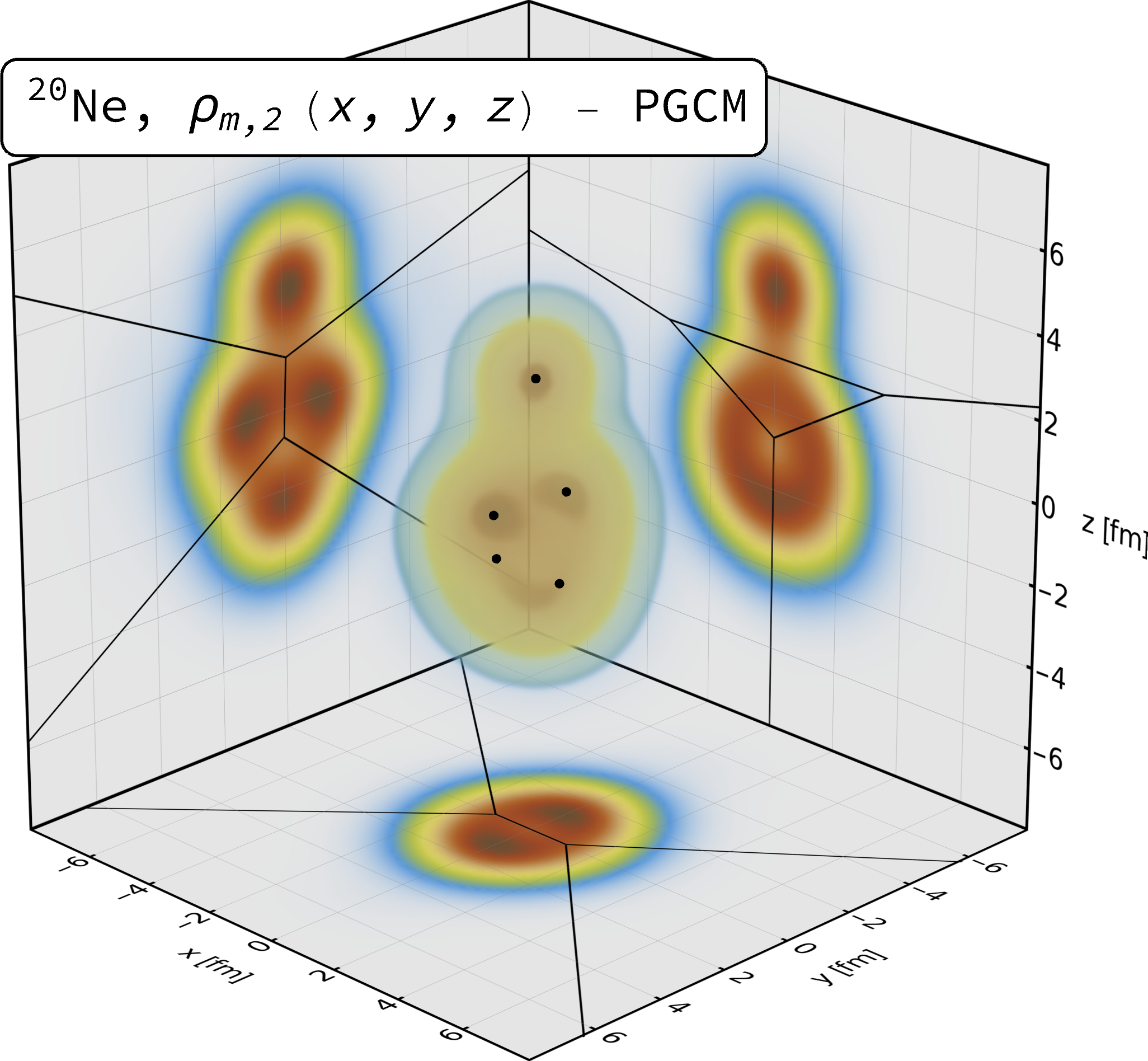}    
    \caption{Point-nucleon densities of $^{16}$O and $^{20}$Ne obtained from particle-number-projected Hartree-Fock-Bogoliubov states with deformations constrained to the predictions of the \textit{ab initio} PGCM framework. The background plots show slices of the densities through the origin. The black dots and lines show the centers and boundaries of the regions used in the clustered sampling method (see text and SM for details)\@.}
    \label{fig:densities_pgcm}
\end{figure}

\begin{figure*}[t]
\centering
\includegraphics[width=.325\textwidth]{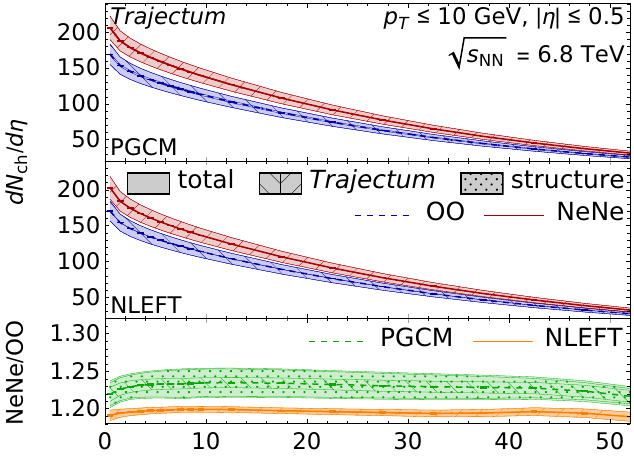}
\includegraphics[width=.325\textwidth]{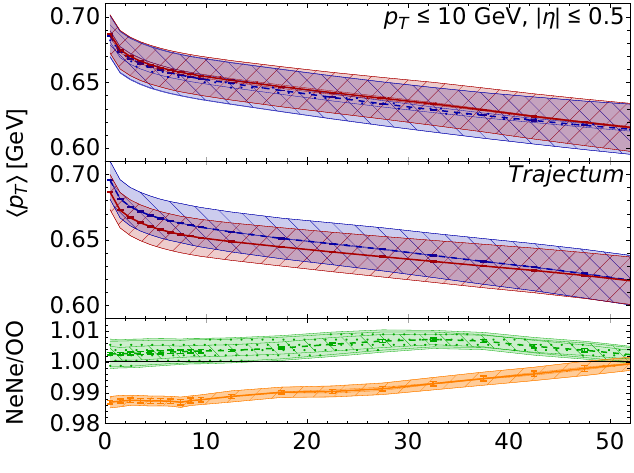}
\includegraphics[width=.325\textwidth]{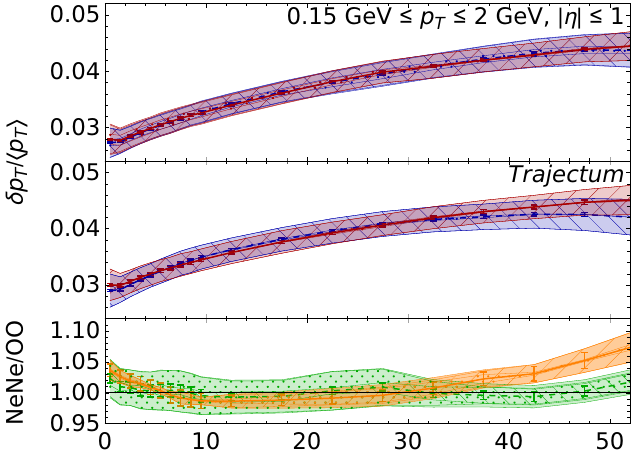}
\includegraphics[width=.325\textwidth]{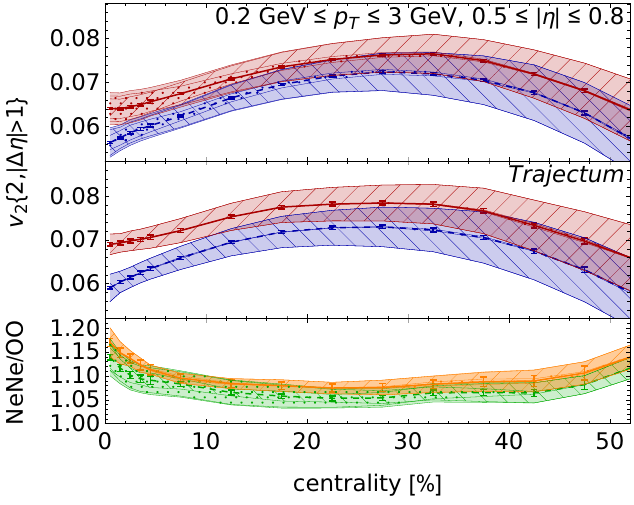}
\includegraphics[width=.325\textwidth]{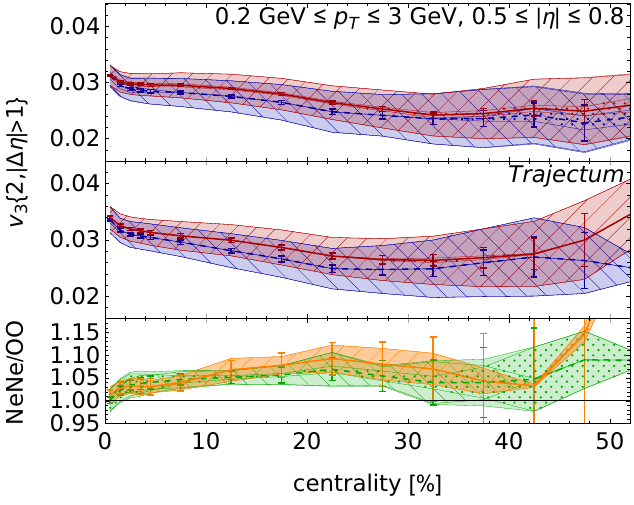}
\includegraphics[width=.325\textwidth]{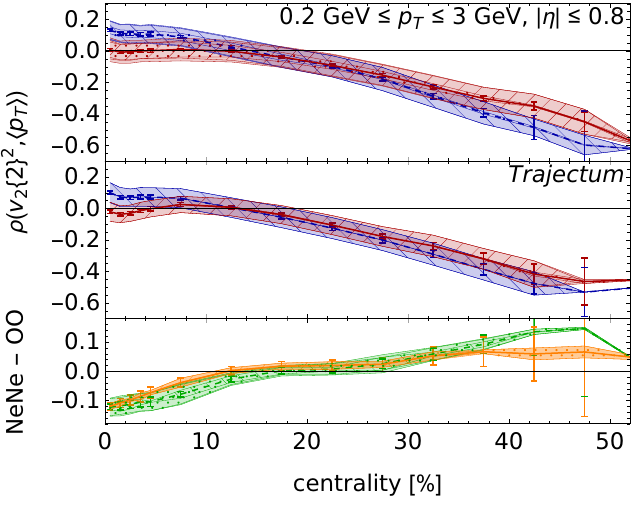}
\caption{The deformed shape of $^{20}$Ne impacts the hydrodynamic flow of its collisions as compared to \oooo{} collisions. Here we show results for charged particle multiplicity $dN_\text{ch}/d\eta$ (top left), mean transverse momentum $\langle p_T\rangle$ (top middle), relative fluctuations of transverse momentum $\delta p_T/\langle p_T\rangle$ (top right), elliptic flow $v_2\{2,|\Delta\eta|>1\}$ (bottom left), triangular flow $v_3\{2,|\Delta\eta|>1\}$ (bottom middle) and the Pearson correlation coefficient $\rho(v_2\{2\}^2,\langle p_T\rangle)$ (bottom right)\@. In each panel, we show the \oooo{} and \nene{} results, as well as their ratio, using both PGCM and NLEFT as nuclear structure inputs. For $\rho(v_2\{2\}^2,\langle p_T\rangle)$ a difference is taken instead of a ratio in the lower panel. We show statistical uncertainties (error bars), the total systematic uncertainty (solid bands) as well as its components being \emph{Trajectum} (hatched) and nuclear structure (dotted)\@.}
\label{fig:2}
\end{figure*}

The \emph{ab initio} PGCM formalism \cite{Frosini:2021tuj,Yao:2018qjv,Yao:2019rck,Frosini:2021fjf,Frosini:2021sxj,Frosini:2021ddm} is also adapted to describe collective correlations, e.g.~quadrupolar and octupolar deformations that appear in doubly-open-shell systems such as $^{20}$Ne. In particular, it was shown in Ref.~\cite{Frosini:2021sxj} that this method captures experimental data on the ground-state rotational band and the charge density of this nucleus, employing a recent N$^3$LO chiral EFT Hamiltonian \cite{Huther:2019ont} which we also use here. We first perform PGCM calculations exploring simultaneously the triaxial quadrupole ($\beta^v_{20}, \beta^v_{22}$) and octupole ($\beta^v_{30}, \beta^v_{32}$) degrees of freedom to determine average intrinsic deformations for the correlated ground states of $^{16}$O and $^{20}$Ne. The resulting shape parameters align with the results of empirical frameworks such as the energy density functional approach~\cite{Marcos:1983msl,Robledo:2011nf,Marevic:2018crl} or the antisymmetrized molecular dynamics approach~\cite{Kimura:2016fce}\@. Then, we compute an intrinsic Hartree-Fock-Bogoliubov state constrained at those average deformations, and we evaluate the particle-number projected one-body density of the resulting system. To quantify the systematic uncertainty on the procedure, the average deformations of the ground states are computed from pure mean-field states as well as from particle-number-projected states (more details in the SM)\@. The results in the latter case are shown in Fig.~\ref{fig:densities_pgcm}\@. We note deformed geometries with well-separated clusters. In $^{16}$O they form an irregular tetrahedron with two short and two long edges of 2.30 and 2.55\,fm respectively (see \cite{YuanyuanWang:2024bxv} for recent work employing a regular tetrahedron)\@. For $^{20}$Ne we observe a characteristic bowling-pin-like $^{16}$O+$\alpha$\@.

For the hydrodynamic simulations, the densities in Fig.~\ref{fig:densities_pgcm} are randomly oriented and used to sample either 16 or 20 coordinates of nucleons for each realization of the nucleus. Unlike the NLEFT simulations, PGCM does not provide us with correlated samplings of nucleon positions. Sampling nucleons capturing the ground-state correlations of the N$^{3}$LO Hamiltonian is therefore ambiguous. We use two methods as a quantification of this systematic uncertainty. One samples nucleons independently (as in \cite{Bally:2021qys,Bally:2022rhf}), whereas the second divides up space into four or five regions (see Fig.~\ref{fig:densities_pgcm}) and samples exactly two protons and two neutrons from each (see also SM)\@. Lastly, configurations are rejected if nucleons are closer than $d_{\rm min}$\@.

\paragraph{Hydrodynamic simulations.} We perform event-by-event hydrodynamic simulations of \nene{} and \oooo{} collisions by means of the \textit{Trajectum} framework \cite{Nijs:2020ors,Nijs:2020roc,Nijs:2021clz,Nijs:2022rme}\@. The calculations start with configurations of nucleons in the colliding nuclei, taken from either the PGCM or the NLEFT results.\footnote{For all profiles we provide 20k configurations as part of the submission.} Each collision is then assigned to an impact parameter, participant nucleons are selected, and energy density is deposited in the transverse plane. Following a brief pre-equilibrium phase, the system is evolved as a relativistic viscous fluid. Hydrodynamic cooling lasts until the local temperature reaches a critical value ($T\sim154$\,MeV), below which hadronization occurs. Subsequent strong decays and rescattering of hadrons are computed by the SMASH code \cite{Weil:2016zrk,dmytro_oliinychenko_2020_3742965,Sjostrand:2007gs}, leading to the particle distributions in the final state. These are analyzed to construct multi-particle correlations following the experimental protocols. We define the collision centrality from the multiplicity of charged particles with $p_T \geq 0.4\,\text{GeV}$ and $|\eta| \leq 2.4$, with 0\% centrality corresponding to the limit of small impact parameters.

The parameters of the model are chosen probabilistically by sampling from the posterior distribution inferred in a Bayesian analysis of \pbpb{} collisions, within the same model \cite{Giacalone:2023cet}\@. We use twenty different samples from the parameter space to quantify the uncertainty on the results coming from wide parameter variations. This represents the largest part of the \emph{Trajectum} systematic uncertainty, which in addition also takes into account effects of finite grid spacing (as discussed in the SM)\@.

Our results for $p_T$-integrated observables that characterize the collective flow of hadrons are displayed in Fig.~\ref{fig:2}\@. Our first remark concerns the cancellation of uncertainties we observe when a relative variation of observables, e.g.~a ratio, is taken between \oooo{} and \nene{} collisions. The dominant uncertainty on the absolute magnitude of the results (upper two plots in each panel) is the systematic one. However, in the relative variations (lowest plots) the contribution from the systematic error becomes nearly equal to that from the statistical error. This enables us to make robust predictions for percent-level variations of observables across the two systems. The larger uncertainty affecting the PGCM results is due to the ambiguities of the empirical method used to extract the correlated distributions of nucleons.

We discuss now those observables that are more strongly impacted by the bowling-pin shape of $^{20}$Ne. The first is the rms elliptic flow, $v_2\{2\}$, in the lower-left panel of Fig.~\ref{fig:2}\@. We find:
\[
\frac{v_2\{2\}_\text{NeNe}}{v_2\{2\}_\text{OO}} = \begin{cases}
1.170(8)_\text{stat.}(30)_\text{syst.}^\textit{Traj.}(0)_\text{syst.}^\text{str.} & \text{(NLEFT)}, \\
1.139(6)_\text{stat.}(27)_\text{syst.}^\textit{Traj.}(28)_\text{syst.}^\text{str.} & \text{(PGCM)},
\end{cases}
\]
in the 0--1\% most central events. This is nearly identical for both nuclear structure inputs, implying that the enhancement of fluctuations in the second harmonic predicted by the NLEFT simulations for \nene{} collisions is largely captured by the (randomly-oriented) bowling pin predicted by the PGCM calculation. The $v_2\{2\}$ ratio between \nene{} and \oooo{} collisions is as large as that expected between peripheral ($\sim$60\% off-central) \pbpb{} collisions and central \oooo{} collisions \cite{Schenke:2020mbo,Nijs:2021clz}\@. However, the cancellation of uncertainties that we achieve here is only possible because we consider experiments with two ions close in mass.

Another probe of the bowling-pin shape of $^{20}$Ne is the correlation between the mean squared elliptic flow, $v_2\{2\}^2$, and the mean transverse momentum, $\langle p_T \rangle$\@. It is quantified via a Pearson coefficient denoted by $\rho_2 \equiv \rho(v_2\{2\}^2,\langle p_T\rangle)$ \cite{Bozek:2016yoj}, which reflects the correlation between the shape and the size of the produced QGP droplets \cite{Bozek:2020drh,Schenke:2020mbo,Giacalone:2020dln}\@. Results for $\rho_2$ are reported in the lower-right panel of Fig.~\ref{fig:2}\@. The suppression of the observable in central \nene{} collisions relative to \oooo{}, observed for both nuclear structure inputs, is a generic signature of the elongated nuclear shape \cite{Giacalone:2019pca,Giacalone:2020awm,Jia:2021wbq,Bally:2021qys,Magdy:2022cvt}\@. The same effect has been reported in \uuuu{} \cite{STAR:2024ekymanual} and \xexe{} \cite{ALICE:2021gxt,ATLAS:2022dov} experiments.

The $\rho_2$ correlator is strongly sensitive to several hydrodynamic model parameters, and thus plagued by a large systematic uncertainty which makes \oooo{} and \nene{} results overlap. Neglecting the triaxiality of these nuclei, and dubbing $\beta_2$ the nuclear quadrupole deformation (where $\beta_{2,\rm ^{20}Ne} > \beta_{2,\rm ^{16}O}$ from spectroscopic data \cite{Pritychenko:2013gwa}, as well as from the densities shown in Fig.~\ref{fig:densities_pgcm}), the $\rho_2$ observable roughly follows at a given centrality: $\rho_2 = a - b \beta_2^3$, where $a$ and $b$ are positive coefficients \cite{Jia:2021qyu, Giacalone:2023hwk, STAR:2024ekymanual}\@. Model studies suggest that both $a$ and $b$ are nearly independent of the collision system at the same centrality \cite{Jia:2021qyu,Bally:2021qys}\@. As a consequence, we expect the difference $\rho_{2,\rm Ne+Ne} - \rho_{2,\rm O+O} \propto \left ( \beta_{2,\rm ^{16}O}^3 - \beta_{2,\rm ^{20}Ne}^3 \right )$ to isolate the imprint of the nuclear deformation. This is confirmed in Fig.~\ref{fig:2} (lower-right panel), where the evaluated difference cancels most of the systematic uncertainties. A comment is in order. In hydrodynamics, the $\rho_2$ of ultra-central \oooo{} collisions is about the same as that of peripheral \pbpb{} collisions at the same multiplicities \cite{Giacalone:2020byk,Nijs:2022rme}\@. Therefore, contrary to the enhancement of $v_2\{2\}$ relative to \oooo{} systems, which occurs in both central \nene{} and peripheral \pbpb{} collisions, the suppression of $\rho_2$ represents a geometry-driven effect only accessible by colliding $^{20}$Ne isotopes.

Four more observables are in Fig.~\ref{fig:2}, namely the charged multiplicity, $dN_{\rm ch}/d\eta$, the mean transverse momentum, $\langle p_T \rangle$, the fluctuations thereof, and the triangular flow, $v_3\{2\}$\@. Significant differences appear between PGCM and NLEFT for $dN_{\rm ch}/d\eta$ and $\langle p_T \rangle$ in the ratio plots. These can be understood from the respective nuclear radii.\footnote{Here the charge radii equal the matter radii for NLEFT since the computation is isospin symmetric. For PGCM the matter radii are about 0.012\,fm smaller than the charge radii.} The NLEFT charge RMS radii are 2.76 and $3.17\,$fm for \oo{} and \nee{} respectively (ratio 1.14), whereas clustered PGCM has 2.87 and $3.09\,$fm with ratio 1.08. For both NLEFT and PGCM we use a Gaussian nucleon charge distribution of width 0.84\,fm \cite{ParticleDataGroup:2020ssz,Lin:2021xrc}\@. This compares well with the experimental values 2.6955 and 3.0055 fm (ratio 1.11) \cite{Angeli:2013epw}\@. We note that for PGCM the independent sampling method gives 0.05 and 0.03\,fm smaller radii for \oo{} and \nee{} respectively. The $d_\text{min}$ parameter has negligible effect when smaller than 0.5\,fm, but increases especially the PGCM radii for larger values. Due to the relatively larger difference in size comparing \nee{} and \oo{}, the NLEFT results lead to a smaller \mpt{} for \nene{} as compared to the PGCM results due to a reduced radial expansion. Similarly, the larger size of the PGCM oxygen leads to an increased \oooo{} cross section and consequently per collision a lower multiplicity, affecting the $dN_{\rm ch}/d\eta$ ratio (see also \cite{Giacalone:2023cet})\@. For the fluctuations of $\langle p_T \rangle$ the observed mild enhancement in central \nene{} collisions is a generic consequence of the more deformed \nee{} shape, which enhances fluctuations in the overall size of the overlap region \cite{Jia:2021qyu,Fortier:2023xxy}\@.

\paragraph{Conclusion \& Outlook.} 

We have showcased the possibility of reducing theoretical systematic uncertainties in hydrodynamic model calculations of small systems. One needs experiments with two light-ion species presenting sufficiently different geometries to perform quantitative tests of the QGP paradigm for shape-induced modifications of the collective flow. As \oooo{} collisions are essentially already available at colliders, the extreme shape of $^{20}$Ne makes this proposal realizable in practice.

Our predictions are based on the same hydrodynamic picture used in the description of collisions of heavy nuclei. They do not include additional ingredients, e.g., out-of-equilibrium corrections due to the expected partial thermalization of the interaction region \cite{Ambrus:2022qya,Kanakubo:2021qcw,Kanakubo:2022ual}, the breakdown of the equations used in our simulations due to causality constraints \cite{Bemfica:2020xym,Plumberg:2021bme,Krupczak:2023jpa}, features such as the escape mechanism \cite{He:2015hfa} that would affect the elliptic flow in a transport-based approach, or even the impact of momentum anisotropies originating in the initial states \cite{Altinoluk:2020wpf}\@. Therefore, testing our predictions in experiments will provide unprecedented quantitative insights into the applicability of a QGP paradigm and the emergence of collective dynamics in QCD matter.

As an outlook, additional research avenues are opened by the study of high-energy \nene{} collisions. They will be the subject of future works.

The elongated $^{20}$Ne shape may help reveal hard-probe modifications in a small system via the study of path-length-dependent effects in the comparison between \nene{} and \oooo{} collisions. These can be studied experimentally by triggering on ultra-central events presenting large final-state ellipticities \cite{Schukraft:2012ah}, and estimated theoretically from the analysis of the path lengths traversed by the hard probes \cite{Beattie:2022ojg}\@.

Second, both $^{16}$O and $^{20}$Ne can be injected in the SMOG2 system of the LHCb detector to perform fixed-target experiments at $\sqrt{s_{\rm NN}}\approx 0.07$\,TeV in the center-of-mass frame \cite{LHCb:2021ysy}\@. The LHC could thus deliver \nene{} and \oooo{} collisions at both $\sqrt{s_{\rm NN}}\approx 7$ and 0.07\,TeV, as well as fixed-target $^{20}$Ne+$^{208}$Pb and $^{16}$O+$^{208}$Pb collisions. This wealth of experimental information combined with the possibility of canceling uncertainties via the study of relative observables would provide a unique handle on the manifestations of small-$x$ dynamics and nonlinear QCD evolution and how they impact the collective structure of nuclei \cite{Giacalone:2023fbe}\@.

Finally, several $\gamma$-mediated processes in ultra-peripheral nucleus-nucleus collisions (UPCs) are aimed at imaging the gluonic content of nuclei at high energy. As showcased in Ref.~\cite{Mantysaari:2023qsq} for the diffractive production of J/$\psi$, the shape of $^{20}$Ne can leave distinct signatures on the cross sections on top of a $\gamma$+$^{16}$O background.  Ratios of observables in UPCs would allow one to cancel uncertainties and obtain a more transparent view of the gluon geometries and their modification at high energy.

\bigskip
\paragraph{Acknowledgements.} We acknowledge the participants of the EMMI Rapid Reaction Task Force \textit{``Nuclear physics confronts relativistic collisions of isobars''} and of the ESNT workshop \textit{``Deciphering nuclear phenomenology across energy scales''} for fruitful discussions about the applications of $^{20}$Ne, which have triggered this project. 
G.G.~is funded by the Deutsche Forschungsgemeinschaft (DFG, German Research Foundation) (German Research Foundation) – Project-ID 273811115 – SFB 1225 ISOQUANT, and under Germany's Excellence Strategy EXC2181/1-390900948 (the Heidelberg STRUCTURES Excellence Cluster).
We also acknowledge support from the U.S. Department of Energy (DE-SC0024586, DE-SC0023658, DE-SC0013365, DE-SC0023175) and the U.S. National Science Foundation (PHY-2310620). 
This work is supported in part by the European
Research Council (ERC) under the European Union's Horizon 2020 research
and innovation programme (ERC AdG EXOTIC, grant agreement No. 101018170),
by DFG and NSFC through funds provided to the
Sino-German CRC 110 ``Symmetries and the Emergence of Structure in QCD" (NSFC
Grant No.~11621131001, DFG Grant No.~TRR110).
The work of UGM was supported in part by the CAS President's International
Fellowship Initiative (PIFI) (Grant No.~2018DM0034).
R.R.~is supported by the Deutsche Forschungsgemeinschaft (DFG, German Research Foundation) – Projektnummer 279384907 – SFB 1245.
T.R.R.~is supported by the Spanish MCIU (PID2021- 127890NB-I00).
The PGCM calculations were performed using HPC resources from GENCI-TGCC (Contracts No.\ A0130513012 and A0150513012) and CCRT (TOPAZE supercomputer). 
The NLEFT code development and production calculations utilized the following computational resources: the Gauss
Centre for Supercomputing e.V. (www.gauss-centre.eu) for computing time on the GCS Supercomputer JUWELS at J{\"u}lich
Supercomputing Centre (JSC) and special GPU time allocated on JURECA-DC;
the Oak Ridge Leadership Computing Facility through the INCITE award
``Ab-initio nuclear structure and nuclear reactions''; and the TUBITAK ULAKBIM High Performance and Grid Computing Center (TRUBA resources).
\bibliographystyle{apsrev4-1}
\bibliography{arXiv,manual}

\newpage

\section{Nuclear structure calculations}

\subsection{NLEFT}

Nuclear Lattice Effective Field Theory (NLEFT) is a method to solve the nuclear $A$-body problem in a finite volume, more precisely on a hypercubic
lattice of volume $L\times L\times L\times L_t$, with $L$ the spatial extent
and $L_t$ the temporal one. It also employs corresponding lattice spacings $a$ and $a_t$\@. Here, the lattice spacing $a$ is typical
chosen in the range from 1 to 2\,fm, corresponding to a UV cutoff of $p_{\rm max} = \pi/a = 314$ to $628$~MeV\@. In this work, we use $a =1.3155$\,fm.

There are various formulations of the underlying action. Here, we employ the so-called minimal nuclear model~\cite{Lu:2018bat}, which has been successfully used to describe the gross properties of light and medium-mass nuclei and the equation of state of neutron matter to a few percent accuracy. It has also been successfully applied to studies of nuclear thermodynamics~\cite{Lu:2019nbg}, studies of clustering in hot dilute matter~\cite{Ren:2023ued} and the puzzling $^4$He transition form factor~\cite{Meissner:2023cvo}\@.
This minimal nuclear interaction is given by the SU(4)-invariant leading-order effective field theory based on smeared contact terms,
\[
H_{{\rm SU(4)}}=H_{\rm free}+\frac{1}{2!}C_{2}\sum_{\mathbf{n}}\tilde{\rho}(\mathbf{n})^{2}
+\frac{1}{3!}C_{3}\sum_{\mathbf{n}}\tilde{\rho}(\mathbf{n})^{3},
\]
where  $\mathbf{n}=(n_{x,}n_{y},n_{z})$ are the lattice coordinates,
$H_{\rm free}$ is the free nucleon Hamiltonian with nucleon mass $m=938.9$\,MeV\@. The density operator $\tilde{\rho}(\mathbf{n})$ is defined as
\cite{Elhatisari:2017eno},
\[
\tilde{\rho}(\mathbf{n})=\sum_{i}\tilde{a}_{i}^{\dagger}(\mathbf{n})\tilde{a}_{i}(\mathbf{n})
+s_{L}\sum_{|\mathbf{n}^{\prime}-\mathbf{n}|=1}\sum_{i}\tilde{a}_{i}^{\dagger}(\mathbf{n}^{\prime})\tilde{a}_{i}(\mathbf{n}^{\prime}),
\]
where $i$ is the joint spin-isospin index, $s_{L}$ is the local smearing parameter, and the non-locally smeared annihilation and
creation operators with parameter $s_{NL}$ are defined as
\[
\tilde{a}_{i}(\mathbf{n})=a_{i}(\mathbf{n})+s_{NL}\sum_{|\mathbf{n}^{\prime}-\mathbf{n}|=1}a_{i}(\mathbf{n}^{\prime}).
\]
The summation over the spin and isospin implies that the interaction is SU(4)
invariant. The parameter $s_L$ ($s_{NL}$) controls the strength of the local (non-local) part of the interaction.  The low-energy constants $C_{2}$ and $C_{3}$ give the overall strength of the two-body
and three-body interactions, respectively.  We take values from~\cite{Lu:2018bat}, $C_2 = -3.41 \cdot 10^{-7}\,$MeV$^{-2}$, 
$C_3 = -1.4 \cdot 10^{-14}\,$MeV$^{-5}$, $s_{NL} = 0.5$, and $s_L = 0.061$\@.

\begin{figure*}[t]
\centering
\includegraphics[width=0.8\columnwidth]{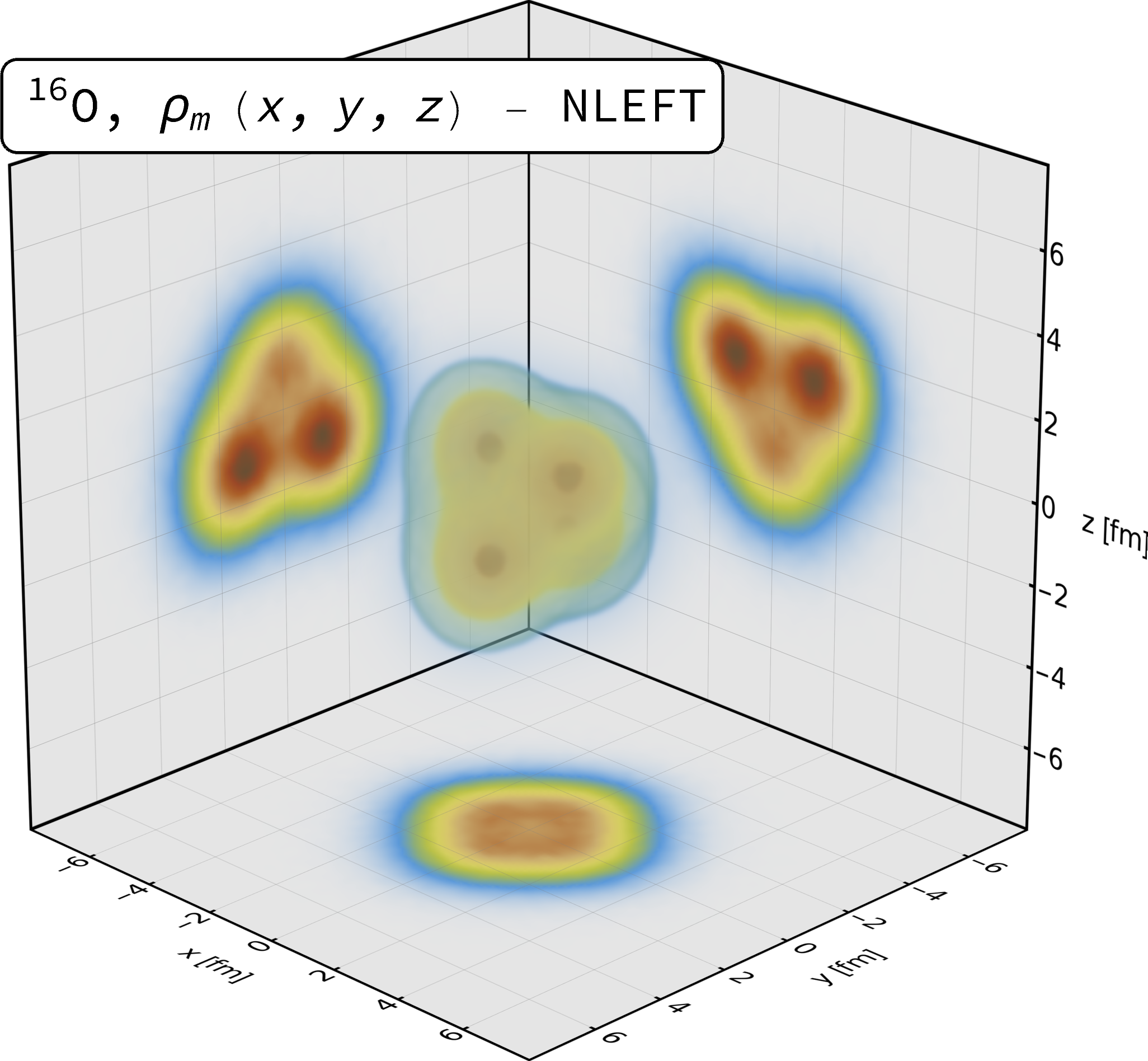}
\hspace{10pt}
\includegraphics[width=0.8\columnwidth]{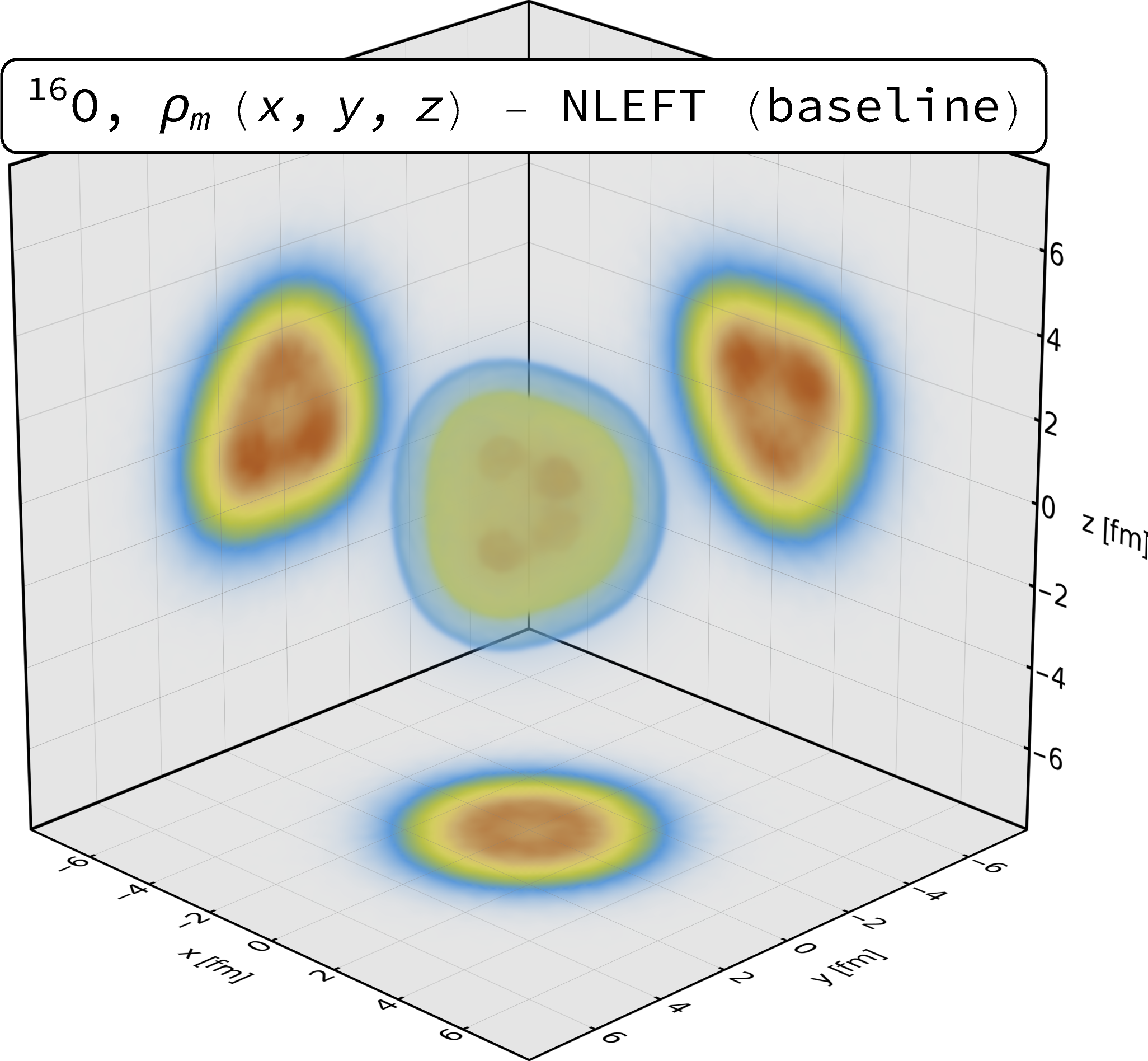}\\
\vspace{10pt}
\includegraphics[width=0.8\columnwidth]{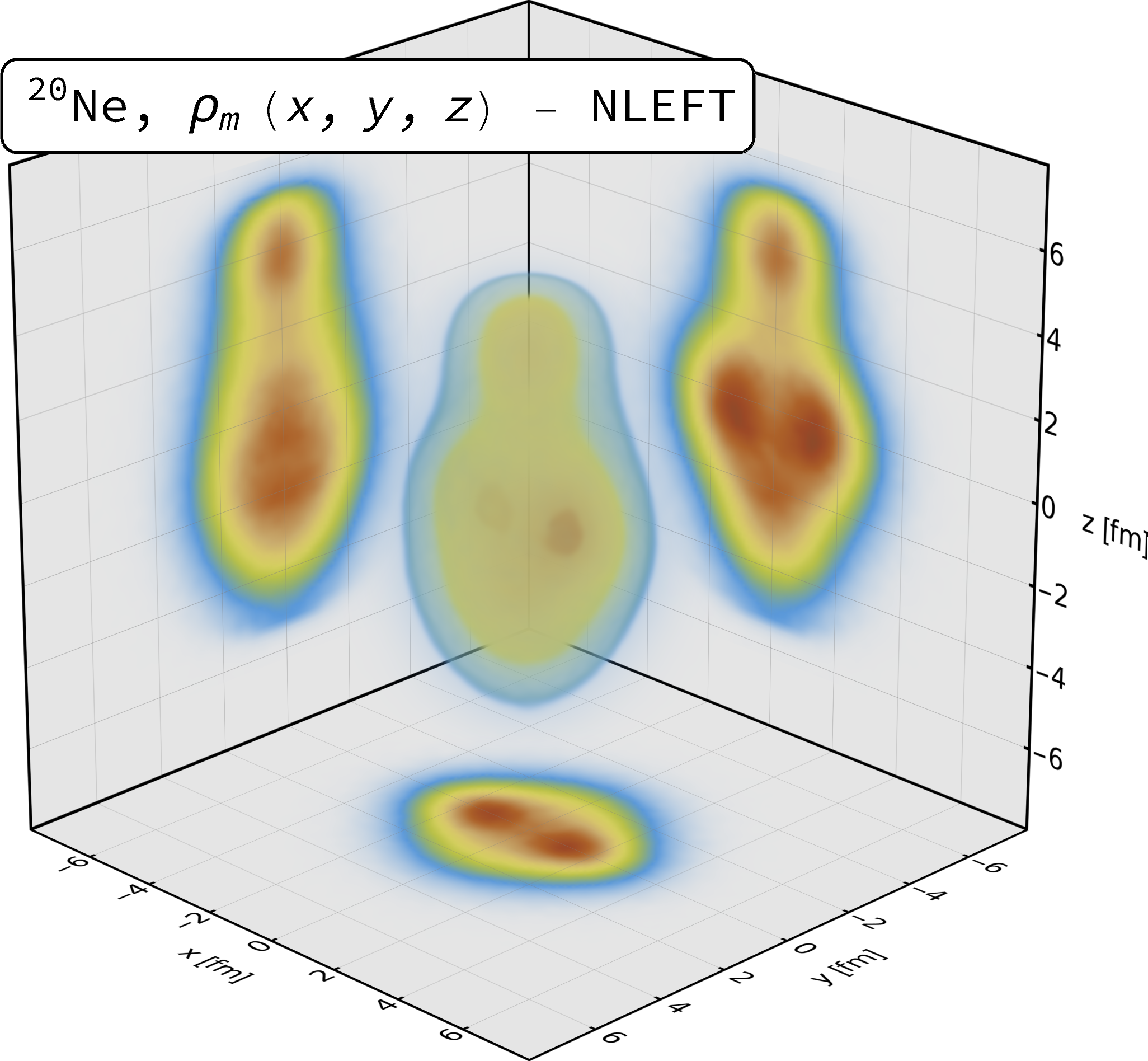}
\hspace{10pt}
\includegraphics[width=0.8\columnwidth]{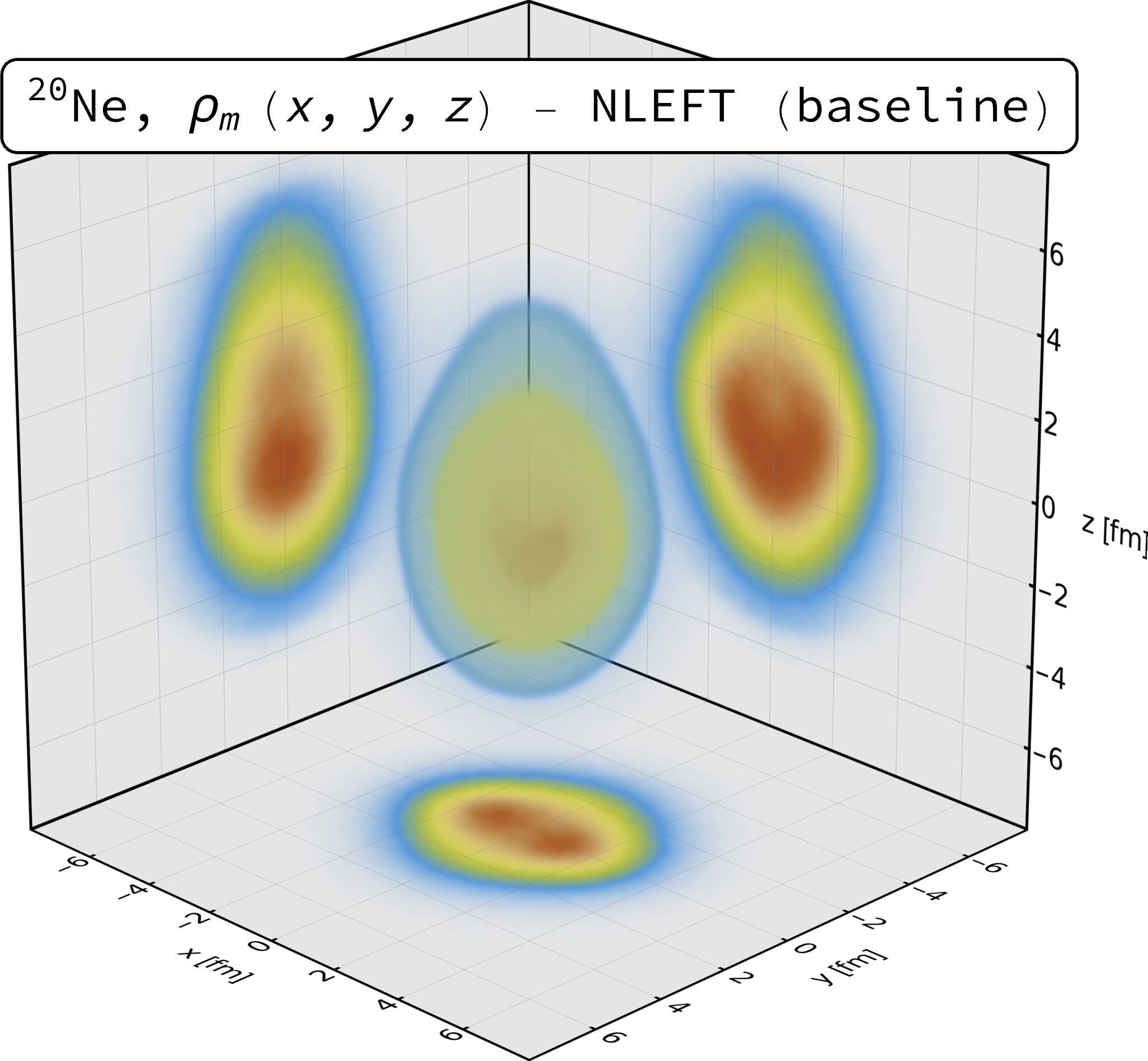}
\caption{Point-nucleon densities of $^{16}$O (top) and $^{20}$Ne (bottom) obtained from NLEFT\@. The panels on the left show densities obtained from configurations aligned as described in the text, whereas the panels on the right show baseline densities obtained from configurations of which the angular information for each nucleon was randomized before the alignment of the nucleus. The background plots show slices of the densities through the origin.}
\label{fig:NLEFT}
\end{figure*}

Central to the studies done here is the pinhole algorithm~\cite{Elhatisari:2017eno}, which performs a Monte Carlo sampling of the $A$-body density of the nucleus in position space. It allows to pin down the center-of-mass of a given nucleus, which is a basic ingredient to calculate charge or matter
distributions as done here. It was e.g.~successfully used to investigate the emergent geometry and intrinsic cluster structure of the low-lying states of $^{12}$C~\cite{Shen:2022bak}\@. We obtain the pinhole algorithm by including the $A$-nucleon density operator
in the projection amplitudes according to
\begin{align*}
Z_{f,i}^{\rm pinhole} & \equiv 
Z_{f,i}^{}(i_1,j_1,\ldots, i_A,j_A; \vec n_1^{}, \ldots, \vec n_A^{}; N_t^{}) \\
& = \langle \Psi_f^{} |  M^{N_t^{}/2}_{}
\rho_{i_1,j_1, \ldots, i_A,j_A}^{}(\vec n_1^{}, \ldots, \vec n_A^{}) \:\\
& \qquad\qquad \times M^{N_t^{}/2}_{}  |\Psi_i^{}\rangle,
\end{align*}
which amounts to the insertion of a ``screen'', with pinholes located at the positions
$\vec n_1, \ldots, \vec n_A$ and spin-isospin indices $i_1,j_1,\ldots, i_A,j_A$, 
at the midpoint of the Euclidean time evolution. Here, the  NLEFT transfer matrix $M$ is applied $N_t$ times  
to the initial and final states. With this, one is able to pin down the center-of-mass of a given nuclear state with 
spin $J$ and parity $P$ and consequently derive charge and matter distributions. It should also be noted that the
resolution of the pinhole algorithm in coordinate space is $a/A$\@.

The pinhole positions contain many-body correlations up to all possible orders.
To reduce them to a one-body intrinsic density distribution, a certain alignment for each configuration is needed, and one should keep in mind that such a procedure may not be unique.
For example, one can align configurations according to the long- or short-axis of the moment of inertia matrix \cite{Wiringa:2000gb,Shen:2022bak}\@.
Here we try to align the pinhole configurations of $^{16}$O and $^{20}$Ne according to the symmetry found by PGCM calculations, i.e., a tetrahedron and bowling pin shape, respectively.
We first identify $\alpha$ clusters by requiring that the distance among the four nucleons (proton spin up/down, and neutron spin up/down) is (approximately) the smallest.
For $^{16}$O, we randomly choose one cluster $\mathbf{c}_1$ and align it to one of the tetrahedron symmetry axis $\mathbf{a}_1$; then we randomly choose another cluster $\mathbf{c}_2$ and symmetry axis $\mathbf{a}_2$, and rotate along $\mathbf{a}_1$ so that the inner product of $\mathbf{c}_2$ and $\mathbf{a}_2$ is the largest.
For $^{20}$Ne, we first align the long principal axis to $z$ direction and require $\sum_{i=1}^A (z_i - z_{\rm c.m.})^3$ to be positive. Then we choose the cluster that has the smallest $|z|$ value to put it in the $\pm x$ directions.

The corresponding point-nucleon densities for the NLEFT calculation are displayed in Fig.~\ref{fig:NLEFT} (left)\@.
The overall shape as well as the inner cluster structure of both nuclei look quite similar to the PGCM results.
The bottom part of $^{20}$Ne looks thinner and the top part looks fatter  than the one obtained by PGCM, this could be due to the effect of aligning the long principal axis to $z-$direction, that the density is strengthened too much along this direction \cite{Wiringa:2000gb}\@.
To study the effect of the alignment procedure, we also show in Fig.~\ref{fig:NLEFT} (right) the same densities, but using configurations where the angular information has been randomized. Without any bias coming from the alignment procedure, these distributions should be exactly spherical. One can therefore see that at least some of the non-trivial shape seen in Fig.~\ref{fig:NLEFT} (left) is due to bias from the alignment procedure. However, the $\alpha$-clustering seen on the left is much more pronounced, as well as the bowling-pin shape of $^{20}$Ne.

\begin{figure*}[t]
\centering
\includegraphics[width=0.49\textwidth]{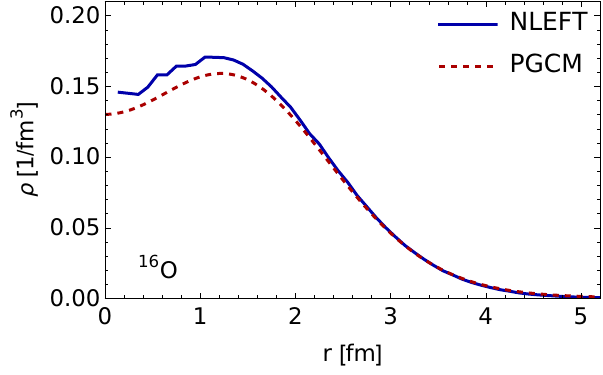}
\includegraphics[width=0.49\textwidth]{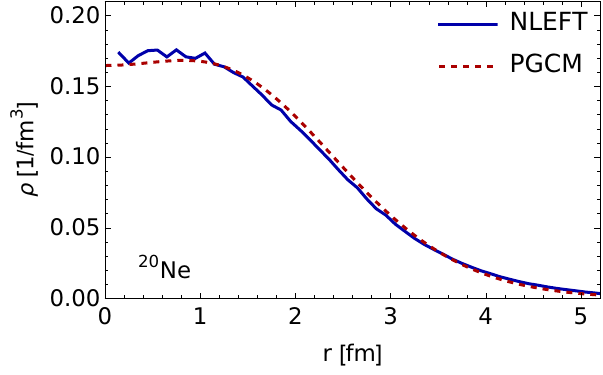}
\caption{\label{fig:nleftpgcmcomparisonplot} We show radial density profiles of $^{16}$O (left) and $^{20}$Ne (right), for both NLEFT and PGCM\@. The PGCM results are computed using the density $\rho_{m,2}$ as given in the text, while the NLEFT results come from smeared lattice configurations with $d_\text{min} = 0.5$\,fm.}
\end{figure*}

\begin{figure*}[t]
\centering
\includegraphics[width=0.325\textwidth]{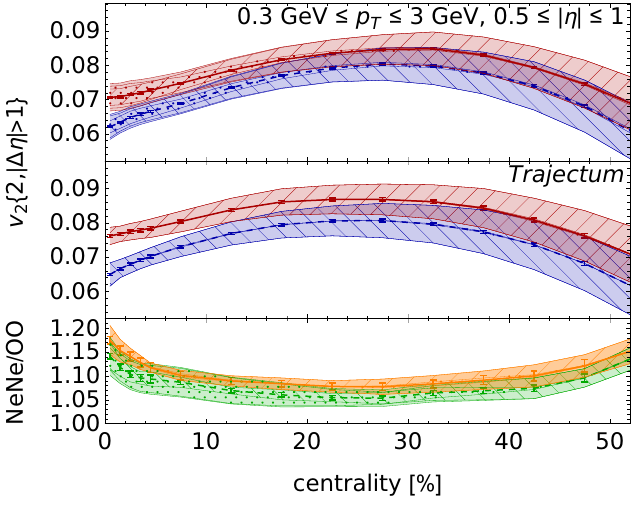}
\includegraphics[width=0.325\textwidth]{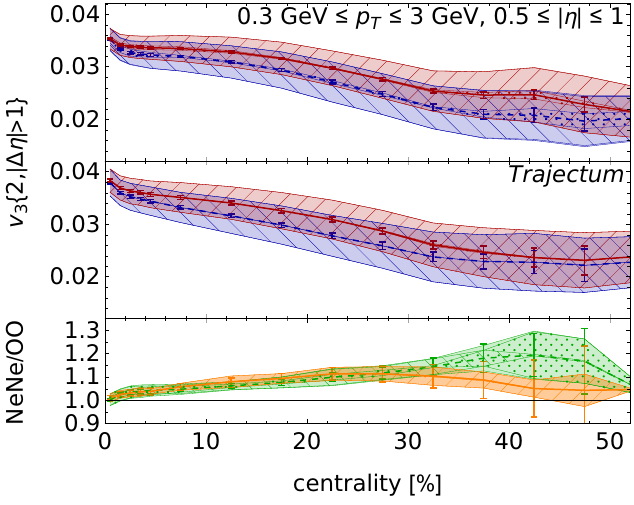}
\includegraphics[width=0.325\textwidth]{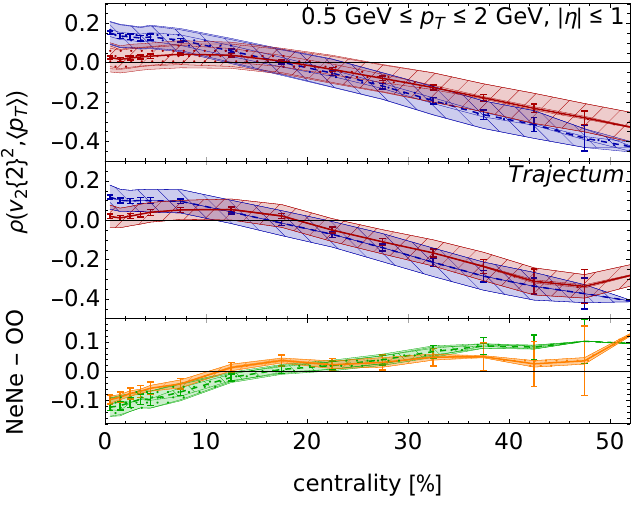}
\caption{\label{fig:altcuts}We show $v_2\{2,|\Delta\eta|>1\}$ (left), $v_3\{2,|\Delta\eta|>1\}$ (middle) and $\rho(v_2\{2\}^2,\langle p_T\rangle)$ (right), with kinematic cuts as used by CMS (left and middle) and ATLAS (right)\@.}
\end{figure*}

\subsection{PGCM}

The calculations described here were performed using the numerical suite TAURUS \cite{Bally:2020kjr,Bally:2024loa} in a model space with $e_\text{max} = 6$, $e_\text{3max} = 18$ and $\hbar \omega = 12$ MeV\@. This model space proved to be sufficient to converge systems with similar masses in previous calculations \cite{Frosini:2021sxj,Bally:2024loa}\@.

Concerning the nuclear Hamiltonian, $H$, we employ the recent $\chi$EFT-based interaction at N$^3$LO published in Ref.~\cite{Huther:2019ont} and apply the rank-reduction scheme proposed in Ref.~\cite{Frosini:2021tuj} to obtain an effective two-body operator.

The first step of our PGCM calculations is the generation of a set of reference states. Here, we consider real general Bogoliubov quasi-particle states optimized at the variation-after-particle-number projection level. In addition, the Bogoliubov states $\ket{\Phi(q)}$ are constrained to have, on average, a given deformation  $q \equiv (\beta_{20}, \beta_{22}, \beta_{30}, \beta_{32})$, where 
\[
\beta_{lm} = \frac{4 \pi}{3 R_0^l A} \langle r^l \left( Y_{lm} + (-1)^m Y_{l-m} \right) \rangle,
\]
with $A$ being the number of nucleons, $R_0 = 1.2 A^{1/3}$, $r$ being the position and $Y_{lm}$ being a spherical harmonic. 
For a more efficient exploration of quadrupole deformations, the constraints on $\beta_{20}$ and $\beta_{22}$ are reformulated as constraints on the usual triaxial parameters
\begin{align*}
\beta &= \sqrt{\beta_{20}^2 + 2 \beta_{22}^2}, \\
\gamma &= \arctan\left( \frac{\sqrt{2} \beta_{22}}{\beta_{20}} \right),
\end{align*}
and we use a parallelogram mesh as described in Ref.~\cite{Rodriguez:2010by}\@.
For $^{16}$O, we considered the intervals:  $\beta \in [0,1.2]$ with a spacing $\Delta \beta = 0.2$, $\gamma \in [0,60^\circ]$, $\beta_{30} \in [0,1.2]$ with a spacing $\Delta \beta_{30} = 0.3$, $\beta_{32} \in [0,1.2]$ with a spacing $\Delta \beta_{32} = 0.3$\@.
For $^{20}$Ne, we considered the intervals:  $\beta \in [0,1.5]$ with a spacing $\Delta \beta = 0.3$, $\gamma \in [0,60^\circ]$, $\beta_{30} \in [0,1.5]$ with a spacing $\Delta \beta_{30} = 0.3$, $\beta_{32} \in [0,0.9]$ with a spacing $\Delta \beta_{32} = 0.3$\@.

To pick the states to be included in the configuration mixing, we project all (converged) quasi-particle states $\ket{\Phi(q)}$ on $J^\pi = 0^+$ and the appropriate number of protons and neutrons, and select the states with a projected energy
\[
E(0^+,q) = \frac{\elma{\Phi(q)}{H P^{\pi=+1} P^{J=0}_{00} P^Z P^N}{\Phi(q)}}{\elma{\Phi(q)}{P^{\pi=+1} P^{J=0}_{00} P^Z P^N}{\Phi(q)}} ,
\]
lower than a threshold of $E_{\text{th}} = 4$ and 8\,MeV for $^{16}$O and $^{20}$Ne, respectively,  above the absolute minimum. Here, $P^{\pi}$, $P^{J}_{MK}$, $P^Z$ and $P^N$ are the projection operators onto a good parity, total angular momentum, number of protons and number of neutrons, respectively \cite{Bally:2020wkb,Sheikh:2020wii}\@. To say it differently, we consider the mixing set 
\[
\Sigma \equiv \left\{ \ket{\Phi(q)} , E(0^+,q) - \min\left[E(0^+,q)\right] < E_{\text{th}} \right\}.
\]

The PGCM ansatz considered for the ground state takes the form
\[
\ket{\Psi^{0^+ ZN}_1} = \sum_q^\Sigma f_{1q}^{0^+ ZN} P^{\pi=+1} P^{J=0}_{00} P^Z P^N \ket{\Phi(q)} ,
\]
with the weights, $f_{1q}^{0^+ ZN}$, being determined variationally by requiring the energy to be minimal. 
After solving the variational equation, we use the weights to determine an ``average'' deformation for the PGCM correlated state, $\ket{\Psi^{0^+ ZN}_1}$, following the procedure described in Ref.~\cite{Bally:2021qys}\@. 
The average deformation is calculated using either the deformation of the Bogoliubov states ($\bar{q}_1$) or the deformation of the associated particle-number projected states ($\bar{q}_2$)\@.
Then, we build a new Bogoliubov state constrained to have the deformation $\bar{q}_1$, $\ket{\Phi(\bar{q}_1})$, as well as a particle-number projected state constrained to have the deformation $\bar{q}_2$, $\ket{\Phi^{ZN}(\bar{q}_2)}$\@.
Finally, we compute the particle-number projected spatial densities 
\begin{align*}
\rho_{m,1} (x,y,z) &= \sum_{st} \frac{\elma{\Phi(\bar{q}_1)}{a^+_{xyzst} a_{xyzst} P^Z P^N}{\Phi(\bar{q}_1)}}{\elma{\Phi(\bar{q}_1)}{ P^Z P^N}{\Phi(\bar{q}_1)}} , \\
\rho_{m,2} (x,y,z) &= \sum_{st} \frac{\elma{\Phi^{ZN}(\bar{q}_2)}{a^+_{xyzst} a_{xyzst}}{\Phi^{ZN}(\bar{q}_2)}}{\scal{\Phi^{ZN}(\bar{q}_2)}{\Phi^{ZN}(\bar{q}_2)}} ,
\end{align*}
where $a^+_{xyzst}$ ($a_{xyzst}$) creates (annihilates) a particle with spin $s=\pm 1/2$ and isospin $t=\pm 1/2$ at position $(x,y,z)$\@.
The two densities are used in the subsequent hydrodynamic simulations to partially estimate the uncertainty related to our empirical procedure.

\subsection{Comparison of configurations}

In Fig.~\ref{fig:nleftpgcmcomparisonplot}, we compare the orientation-averaged radial distributions of nuclei generated using both NLEFT and PGCM\@. For PGCM, we show the density $\rho_{m,2}$ as described above, and for NLEFT we show densities reconstructed from the lattice configurations after they have been smeared and after a minimal distance $d_\text{min} = 0.5$\,fm has been imposed. We take the origin to lie at the center of mass of either the density (PGCM) or at the center of mass of the sampled nucleons (NLEFT)\@. Overall agreement between the two calculations is good, especially for $^{20}$Ne.

\section{Kinematic cuts}

In Fig.~\ref{fig:altcuts}, we show $v_2\{2,|\Delta\eta|>1\}$, $v_3\{2,|\Delta\eta|>1\}$ and $\rho(v_2\{2\}^2,\langle p_T\rangle)$ with different kinematic cuts from those used in the main text. In the main text, cuts as used by ALICE were used, whereas here we use CMS cuts for $v_n\{2,|\Delta\eta|>1\}$, and ATLAS cuts for $\rho(v_2\{2\}^2,\langle p_T\rangle)$\@. One can see that while the individual curves for $^{20}$Ne and $^{16}$O have a dependence on the cuts, the ratios/differences shown in the bottom panel for each observable are unchanged.

\section{Statistical and systematic uncertainties}

\begin{figure}[t]
\centering
\includegraphics[width=\columnwidth]{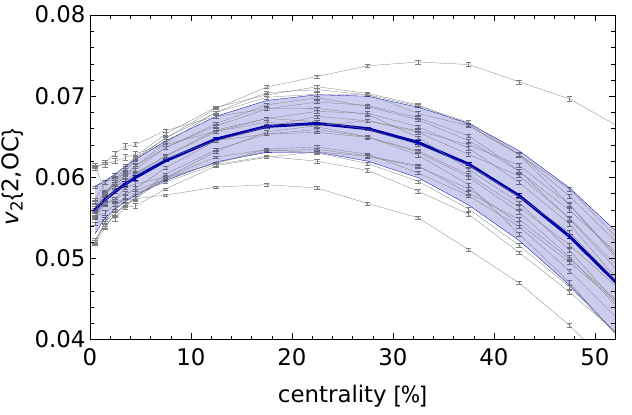}
\caption{\label{fig:inputfileerrorplot}We show $v_2\{2,\text{OC}\}$ for \oooo{} collisions, using NLEFT nuclear structure input. The gray curves represent 20 different parameter choices taken from a Bayesian analysis \cite{Giacalone:2023cet}\@. The blue curve is the mean of the gray curves, and the error bars shown on it represent the statistical uncertainty. The blue band represents the systematic uncertainty, and is constructed from the spread of the gray curves.}
\end{figure}

The observables in the main text are shown with statistical and systematic uncertainties, where the latter are broken up into \emph{Trajectum} and nuclear structure parts. In this section, we discuss in detail how these uncertainties are computed.

To quantify the systematic uncertainty due to the remaining uncertainty in the parameters of the \emph{Trajectum} model after the Bayesian fit, we take 20 likely choices of parameters, which are randomly drawn from our posterior. We subsequently evaluate our observables on all of these choices, and use the spread of the parameters to generate a systematic uncertainty. Since this essentially requires doing the computation 20 times, we choose to do this computation using settings which evaluate relatively quickly. This requires some approximations, which we then subsequently identify and correct for. In this way, the computation is naturally split up into a `main' computation where we use the 20 choices of parameters to generate an ensemble of predictions, and a number of corrections. This section is organized in the same way.

\subsection{\emph{Trajectum}}

First, let us discuss how the ensemble of 20 predictions is used. As an example, in Fig.~\ref{fig:inputfileerrorplot}, we show in gray the 20 predictions in \oooo{} (NLEFT) for $v_2\{2,\text{OC}\}$, an observable related to $v_2\{2,|\Delta\eta|>1\}$ (more on this below)\@. Each prediction comes with its own statistical uncertainty as estimated by the \emph{Trajectum} code. We then average the values to produce the mean value (shown as a blue curve), which we take as our central value. The statistical uncertainty is estimated by averaging the statistical variances of the 20 predictions.\footnote{We average the variances instead of the standard deviations because the variance is an additive quantity.} We then compute the standard deviation of the ensemble of predictions, which gives us the total uncertainty. The systematic uncertainty is then computed as the quadratic difference between the total and the statistical uncertainties. The systematic uncertainty coming from this source is counted as part of the \emph{Trajectum} systematic uncertainty.

\begin{figure}[t]
\centering
\includegraphics[width=\columnwidth]{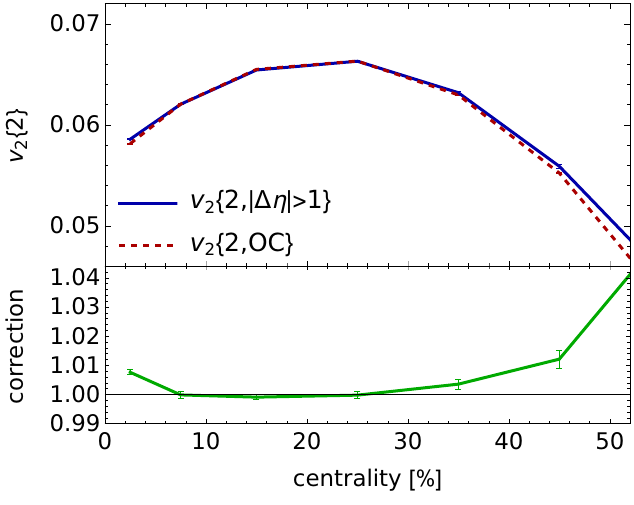}
\caption{\label{fig:obscorrectionplot}We show both $v_2\{2,|\Delta\eta|>1\}$ and $v_2\{2,\text{OC}\}$ for \oooo{} (NLEFT) (top)\@. Also shown is the ratio $v_2\{2,|\Delta\eta|>1\}/v_2\{2,\text{OC}\}$ (bottom), which serves as the correction factor discussed in the text. It can be seen that the correction factor is small but significant.}
\end{figure}

We now turn to a discussion of the corrections, where we start with corrections common to the PGCM and NLEFT simulations. Firstly, we point out an important distinction between the $\rho(v_2\{2\}^2,\langle p_T\rangle)$ observable and the others. Since $\rho(v_2\{2\}^2,\langle p_T\rangle)$ can and does cross zero, we treat its corrections additively, whereas the other corrections are treated multiplicatively. In the discussion that follows, we will be using multiplicative corrections as an example, which will always involve a division of some sorts. The additive corrections work the same way, but replacing division by subtraction.

The first correction we encounter occurs only for $v_2\{2,|\Delta\eta|>1\}$ and $v_3\{2,|\Delta\eta|>1|\}$\@. These observables are statistically hard due to their large $\eta$ gap, but there is a cheaper alternative, which we have called $v_2\{2,\text{OC}\}$, where the OC stands for oversampled cumulants. The observable $v_n\{2\}$ is defined by
\[
v_n\{2\}^2 = \langle\langle\exp(in(\varphi_i-\varphi_j))\rangle\rangle,
\]
where the double $\langle\cdot\rangle$ means first an average over particle pairs $(i,j)$ inside each event, and then an average over events. Which particle pairs are used then determines the subtle differences between various related observables. For $v_n\{2\}$, one simply uses all pairs within the kinematic cuts, whereas for $v_n\{2,|\Delta\eta|>1\}$ one uses only pairs where one particle has $\eta < -0.5$ and the other has $\eta > 0.5$\@. The latter choice suppresses non-flow by using only particle pairs which are far apart in rapidity. Since \emph{Trajectum} is a theory model, we can do one thing which in experiment is impossible; we can generate multiple particle sets from a single hydrodynamic evolution, which we call oversamples. If one analyzes these oversamples separately in the same way as described above, then using oversamples simply improves statistics without introducing a bias. What we can also do, however, is to use the oversamples as a cheap way to mimic the effect of having a (large) $\eta$ gap, while still using all available pairs from an event. The way the observable $v_n\{2,\text{OC}\}$ does this is by letting go of the requirement that the partners in a particle pair need to be from the same oversample. This dilutes the effect of non-flow (in particular resonance decays), thereby mimicking the $v_n\{2,|\Delta\eta|>1\}$ observable, but with much better statistical uncertainty.

\begin{figure}[t]
\centering
\includegraphics[width=\columnwidth]{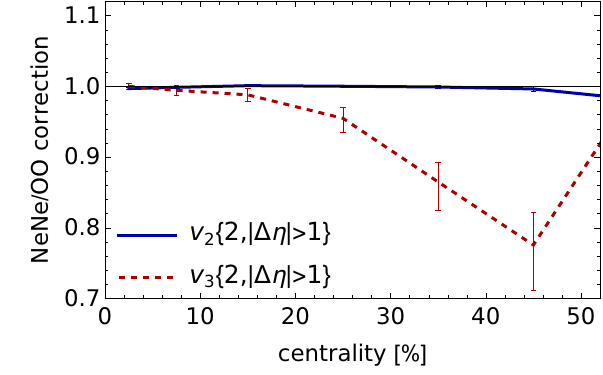}
\caption{\label{fig:obscorrectionratioplot}We show the same correction factor as in the bottom panel of Fig.~\ref{fig:obscorrectionplot}, but for the ratio \nene{}/\oooo{} (NLEFT)\@. We show both the correction for $v_2\{2,|\Delta\eta|>1\}$ and for $v_3\{2,|\Delta\eta|>1\}$\@. It can be seen that whereas the former is a relatively small correction, the latter can be as much as 20\%\@.}
\end{figure}

Returning to the same example used in Fig.~\ref{fig:inputfileerrorplot}, in Fig.~\ref{fig:obscorrectionplot} we show both $v_2\{2,|\Delta\eta|>1\}$ and $v_2\{2,\text{OC}\}$ for \oooo{} using NLEFT, as well as the ratio between both observables. This ratio represents the correction factor with which one needs to multiply the $v_2\{2,\text{OC}\}$ result to obtain $v_2\{2,|\Delta\eta|>1\}$\@. We treat the uncertainties in this correction factor as statistical. Note that whereas for the \nene{}/\oooo{} ratio of $v_2\{2,|\Delta\eta|>1\}$ this correction factor is well under 1\% for most centralities, the same ratio for $v_3\{2,|\Delta\eta|>1\}$ shows a correction of up to 20\% (see Fig.~\ref{fig:obscorrectionratioplot})\@.

\begin{figure}[t]
\centering
\includegraphics[width=\columnwidth]{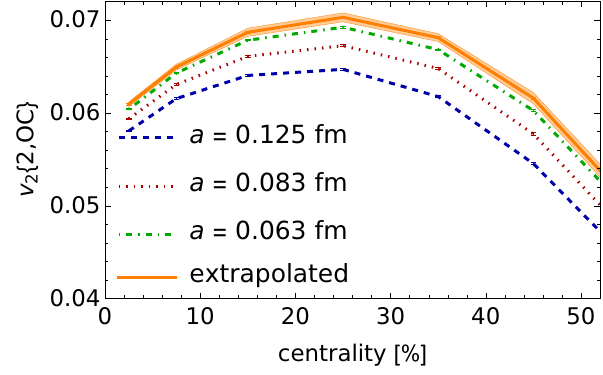}
\caption{\label{fig:extrapolationplot}We show $v_2\{2,\text{OC}\}$ for \oooo{} (NLEFT), using three different grid spacings for the hydrodynamic calculations. In addition, we show the extrapolation to zero grid spacing, where the error bars show statistical uncertainty, and the bands show systematic uncertainty.}
\end{figure}

There is one other correction applied to both PGCM and NLEFT results, and that is a finite grid spacing correction. The hydrodynamic simulations are performed at a finite grid spacing, which can bias the results. To correct for this, we perform calculations using 0.125, 0.083 and 0.063\,fm grid spacings. Continuing to use our example of $v_2\{2,|\Delta\eta|>1\}$, we show these different computations for $v_2\{2,\text{OC}\}$ of \oooo{} for NLEFT in Fig.~\ref{fig:extrapolationplot}\@. We then take a continuum limit by performing a least squares fit of these results to
\[
v_2\{2,\text{OC}\} = c_1 + c_2a^2,
\]
where $c_i$ are constants, and $a$ is the grid spacing. The constant $c_1$ then contains our extrapolated result, which is also shown in Fig.~\ref{fig:extrapolationplot}\@. Using $c_1$ and $c_2$ together, we can then obtain a factor which translates from the grid spacing that the ensemble of 20 predictions was made with to the extrapolated value. We use the prior uncertainties from the fit as statistical uncertainties on the correction factor, and if the posterior uncertainties are larger (which happens if $(\chi^2)_\text{red} > 1$), then we use the squared difference of the posterior and prior uncertainties as a systematic, which we count as part of the \emph{Trajectum} systematic uncertainty.

\subsection{NLEFT}

\begin{figure}[t]
\centering
\includegraphics[width=\columnwidth]{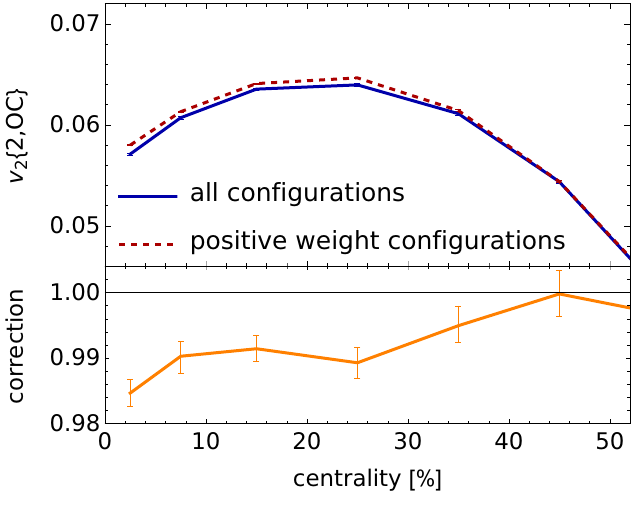}
\caption{\label{fig:weightcorrectionplot}We show $v_2\{2,\text{OC}\}$ for \oooo{} (NLEFT), where we either take only configurations with positive weight, or all configurations. In the latter case, we propagate the weights through the entire computation. The resulting correction factor is shown in the bottom panel.}
\end{figure}

We now describe the corrections which are applied only to the NLEFT results. The first of these pertains to the weights of the NLEFT configurations. For $^{16}$O ($^{20}$Ne), around 26\% (33\%) of configurations have a weight of $-1$, whereas the rest has weight $+1$\@. To take this into account properly, we must assign each collision weight $+1$ if the weights of the two nuclei are both positive or both negative, and $-1$ otherwise. The number of collisions with weight $+1$ is therefore only 62\% (56\%), and this makes the simulations much more expensive statistically. For this reason, we compute the ensemble of 20 predictions using only the configurations with positive weight, and then perform a correction using a dedicated high statistics run using all configurations. In Fig.~\ref{fig:weightcorrectionplot}, we show the results for $v_2\{2,\text{OC}\}$ for \oooo{} using either only the positive weight configurations or all configurations, together with the ratio between the two results, which is used as the correction factor. We then count the statistical uncertainty of this ratio as part of the total statistical uncertainty. We note that the weighting effect is in fact rather important for e.g.~the charge radius. Indeed for \nee{} the positive and negative weight radii are 3.205 and 3.244\,fm respectively, which then leads to an overall radius of 3.17\,fm quoted in the main text.

\begin{figure}[t]
\centering
\includegraphics[width=\columnwidth]{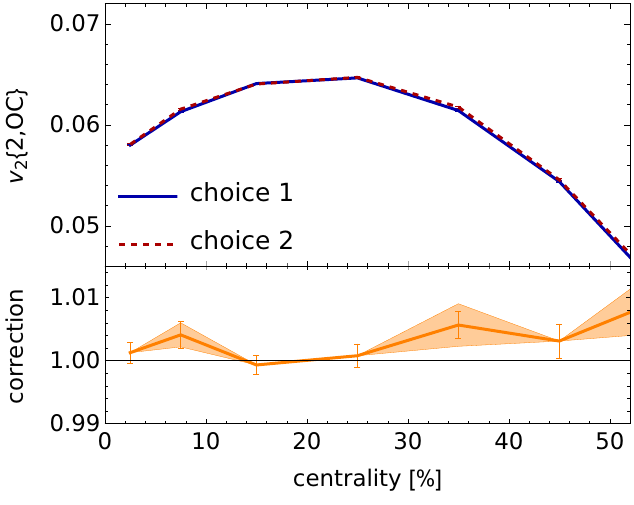}
\caption{\label{fig:ambiguitycorrectionplot}We show $v_2\{2,\text{OC}\}$ for \oooo{} (NLEFT), where we resolve the ambiguity with both choices described in the text. The resulting correction factor is also shown in the bottom panel, where the error bars show statistical uncertainty, and the bands show systematic uncertainty.}
\end{figure}

The other correction applied to NLEFT only deals with resolving the periodicity induced by the lattice. We obtain the configurations on non-periodic boundaries by examining all possible places to put the boundaries between adjacent unit cells, and selecting the choice which leads to the smallest nuclear rms size. In a small number of cases (1.1\% for $^{16}$O and 6.3\% for $^{20}$Ne), this is ambiguous, as there is more than one solution which minimizes the rms size. In this case, we have two choices for resolving this further. The first choice is to just pick randomly, whereas for the second choice we pick such that after clustering into $\alpha$-clusters, the clusters have a smaller size. In other words, the second choice favors $\alpha$-clustering. Given that the unambiguous configurations appear to be $\alpha$-clustered, we believe that the second choice should be better. However, since this is a choice we impose and not something that comes from the Hamiltonian, we view this as a source of systematic uncertainty. We show both methods for $v_2\{2,\text{OC}\}$ in \oooo{} in Fig.~\ref{fig:ambiguitycorrectionplot}, as well as the ratio of choice 2 over choice 1\@. This ratio is used as a correction factor, as the ensemble of 20 predictions was made with choice 1, whereas we believe that choice 2 is better. The standard deviation between the two choices divided by choice 1 is used as a systematic uncertainty after quadratically subtracting the statistical uncertainty of the ratio. This systematic uncertainty is counted as the sole part of the nuclear structure part of the total systematic uncertainty for NLEFT\@. From Fig.~\ref{fig:2} one can see that this uncertainty is almost everywhere negligible compared to the \emph{Trajectum} systematic uncertainty.

\begin{figure*}[t]
\centering
\includegraphics[width=0.49\textwidth]{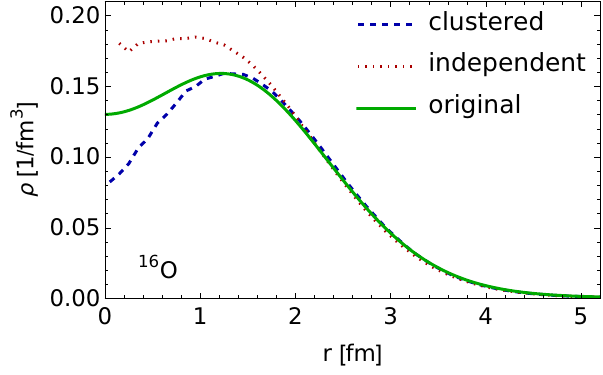}
\includegraphics[width=0.49\textwidth]{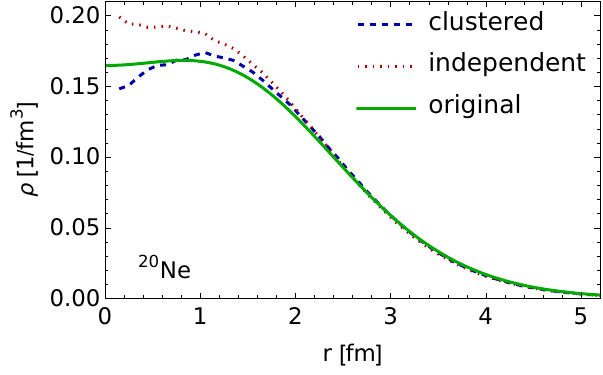}
\caption{\label{fig:clusteringcomparisonplot}We show the radial profiles of $^{16}$O (left) and $^{20}$Ne (right) for PGCM, where we show both sampling methods discussed in the text, alongside the original densities these configurations were sampled from.}
\end{figure*}

\subsection{PGCM}

\begin{figure}[t]
\centering
\includegraphics[width=\columnwidth]{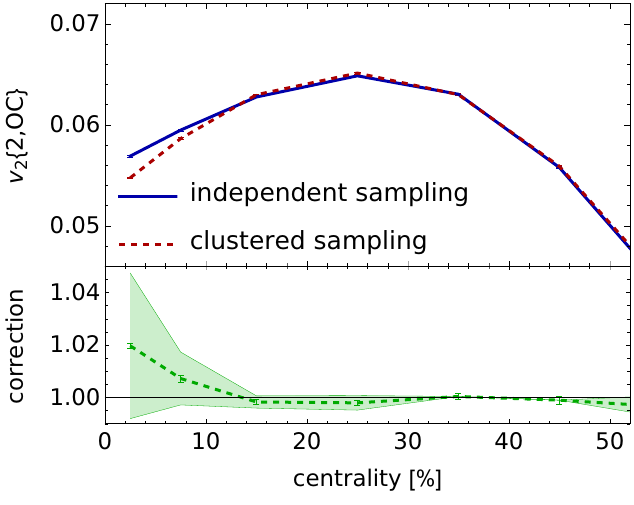}
\caption{\label{fig:clustercorrectionplot}We show $v_2\{2,\text{OC}\}$ for \oooo{} (PGCM), where we either sample independently from the densities provided by PGCM, or we sample such that we enforce that each region (corresponding with an alpha cluster) contains exactly 4 nucleons. The resulting correction factor is shown in the bottom panel.}
\end{figure}

We then move on to describing the two sources of nuclear structure systematic uncertainty for the PGCM results. The first of these sources comes from the fact that PGCM produces a density function, whereas what \emph{Trajectum} needs as its input are explicit configurations. Taking a finite number of samples from a density invariably alters the shape of the sampled distributions compared to the density they were sampled from. As an extreme example of this, when sampling a single nucleon from any density, the resulting shape will always be spherical. Given that the entire effect we study in this work relies on precisely knowing the shapes of the colliding nuclei, it is important to quantify how the shape distortion from sampling affects the final results. To this end, we use two different sampling methods. The first (independent sampling) simply samples 16 (20) independent nucleons from the density. The second (clustered sampling) divides the space up into 4 (5) regions centered on the local maxima defined by the $\alpha$-clusters, where the region boundaries are determined by equidistance between region centers.\footnote{We move the centers slightly compared to these initial ans\"atze to ensure each region has approximately equal integrated density inside.} These regions can be seen in Fig.~\ref{fig:densities_pgcm}, where the centers are shown as black points on the 3D view, and the boundaries between the regions are shown in the 2D cross sections. We then sample 4 nucleons from each region according to the density. The clustered sampling method enforces a more even distribution of sampled nucleons throughout the density, suppressing fluctuations where one of the regions contains more nucleons than another. For this reason, we believe that clustered sampling should yield configurations closer in shape to the original density than independent sampling and importantly is more realistic in approximating the physical sampling where all nucleon-nucleon correlations would have been taken into account. Fig.~\ref{fig:clusteringcomparisonplot} shows the radial profile of both $^{16}$O (left) and $^{20}$Ne (right) using both of these methods, and compares against the original density they were sampled from (shown as well in Fig.~\ref{fig:nleftpgcmcomparisonplot})\@. For $^{16}$O, the clustered method is near the original density from around 1\,fm outwards, whereas for the independent method this is about 1.5\,fm. Below those values the clustered method underestimates the original density, whereas the independent method overestimates it. For $^{20}$Ne, the difference between both methods is more pronounced, with the clustered method being closer to the original density than the independent method almost everywhere.

Given that both methods are approximations, we take an agnostic approach, and use their  mean  as our central value, whereas we use the standard deviation between them as part of the nuclear structure systematic uncertainty (after quadratically subtracting the statistical uncertainty)\@. The ensemble of 20 predictions is made using clustered sampling, by which we normalize both the central value and the systematic uncertainty to obtain the correction factor. Fig.~\ref{fig:clustercorrectionplot} shows the $v_2\{2,\text{OC}\}$ of \oooo{} for both of the sampling methods, as well as the correction factor obtained from them.

\begin{figure}[t]
\centering
\includegraphics[width=\columnwidth]{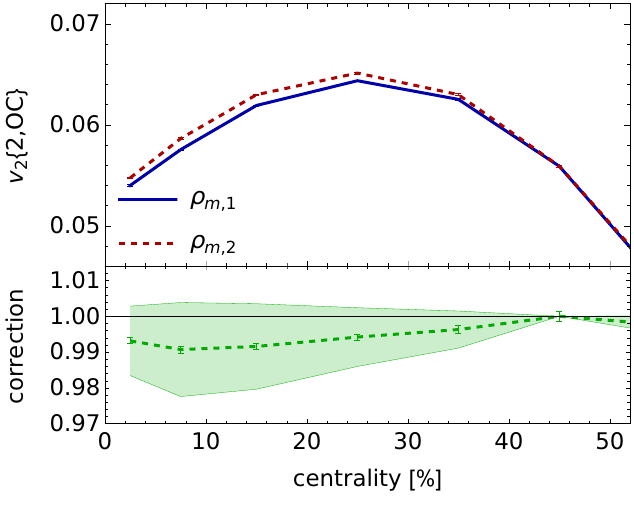}
\caption{\label{fig:projectioncorrectionplot}We show $v_2\{2,\text{OC}\}$ for \oooo{} (PGCM), where we use either the density $\rho_{m,1}$ or $\rho_{m,2}$ as inputs to \emph{Trajectum}\@. The resulting correction factor is shown in the bottom panel.}
\end{figure}

The last source of nuclear structure systematic uncertainty for PGCM is constructed from the two previously discussed methods to constrain the configurations to the ground state, resulting in the two densities $\rho_{m,1}(x,y,z)$ and $\rho_{m,2}(x,y,z)$\@. The results obtained from these two densities give us a handle on the effect of these variations in the PGCM computation on the final state observables. Therefore, similarly to the previous source of systematic uncertainty, we use the mean of the results obtained from both densities as our central value, and use the standard deviation between them as part of the nuclear structure systematic error. As the computation using $\rho_{m,2}(x,y,z)$ is used for the ensemble of 20 predictions, we use it to normalize both the central value and the systematic error. Fig.~\ref{fig:projectioncorrectionplot} shows the $v_2\{2,\text{OC}\}$ of \oooo{} for both of these versions, as well as the correction factor obtained from them.

\end{document}